\documentclass[superscriptaddress,amsmath,amssymb,aps,onecolumn,pra]{revtex4-2}

\usepackage{amssymb}
\usepackage{amsmath}
\usepackage{color}
\usepackage{soul}
\usepackage[utf8]{inputenc} 
\usepackage[T1]{fontenc}    
\usepackage{graphicx}       
\usepackage[margin=2.5cm]{geometry}
\usepackage{float}

\begin{document}


\title{Confrontation of capitalism and socialism in Wikipedia networks}

\author{Leonardo Ermann}
\affiliation{Departamento de F\'{\i}sica Te\'orica, GIyA,
         Comisi\'on Nacional de Energ\'{\i}a At\'omica.
           Av.~del Libertador 8250, 1429 Buenos Aires, Argentina}

\author{Dima L. Shepelyansky}
\email{dima@irsamc.ups-tlse.fr}
\affiliation{Univ Toulouse, CNRS, Laboratoire de Physique Th\'eorique,  
Toulouse, France}

\date{July 31, 2024} 

\begin{abstract}
We introduce the Ising Network Opinion Formation (INOF) model
and apply it for the analysis of networks of 6 Wikipedia language editions.
In the model, Ising spins are placed at network nodes/articles
and the steady-state opinion polarization of spins is determined from
the Monte Carlo iterations in which a given spin orientation is determined by
in-going links from other spins. The main consideration is done
for opinion confrontation between {\it capitalism, imperialism} (blue opinion) and
{\it socialism, communism} (red opinion). These nodes have fixed spin/opinion orientation
while other nodes achieve their steady-state opinions in the process of Monte Carlo iterations.
We find that the global network
opinion favors {\it socialism, communism} for all 6 editions. The model also
determines the opinion preferences for world countries and political leaders,
showing good agreement with heuristic expectations. We also
present results for opinion competition between {\it Christianity} and {\it Islam},
and USA Democratic and Republican parties. We argue that the INOF approach
can find numerous applications for directed complex networks.
\end{abstract}
  



\maketitle

\section{Introduction}

The emergence of social networks,
characterized by scale-free properties (see e.g. \cite{fortunato09,dorogovtsev10}),
produced an important impact on human society. Thus opinion formation
in such social media is argued to influence even political elections
(see e.g. \cite{soc1,soc2}). This implies that the understanding of opinion formation on
social networks represents an important challenge.
Various  voter models on networks
had been developed for the analysis of opinion formation as described in
\cite{fortunato09,sznajd00,sood05,watts07,galam08,schmittmann10,kandiah12,eom15}.
Recently the opinion formation on the world trade network  has been argued to
be linked with country preference to trade in one or another currency
(e.g. US dollar or hypothetical BRICS currency) \cite{brics2024}.
An important new element appeared in these studies is that opinion
of certain countries (network nodes) is considered to be fixed since
it is assumed that they prefer to trade always with fixed currency of
USD or BRICS. This raises the question of how important is an influence of
specific selected nodes with opposite opinions on a global opinion configuration
in  complex directed networks.

A network with $N$ nodes and two opinions
can be viewed as a generalized Ising model with spins $\sigma= \pm 1$.
The total number of opinion (or spin) configurations in such a system
is huge being $N_s =2^N$.
It is natural to assume that a given voter, or node, opinion is 
determined by the opinions of directly linked neighbors
that makes the problem to be similar to a spin
polarization (or magnetization) in the Ising model:
if the  neighboring spins of a specific spin, or a voter, are mainly
up-oriented (red color) then this spin also turns up,
or if the neighboring spins are down-oriented (blue color),
then the spin turns down. Such an approach to opinion formation on various
networks had been applied in many cases and analyzed in the above cited publications.

In this work we study the problem of opinion formation
induced by a group of nodes with fixed polarization (opinion)
in the Wikipedia networks of different languages
(up to 6 ones, with English-EN, German-DE,  Spanish-ES, French-FR, Italian-IT, Russian-RU).
We use the network data sets of Wikipedia collected in 2017
and publicly available at \cite{wiki2017}.
The important advantage of Wikipedia networks
is that the meaning of its nodes is
well clear from the corresponding articles of Wikipedia.
A number of features of WIKI2017 networks had been
studied e.g. in \cite{wrwu2017,wikibanks}.
A great  variety of applications of Wikipedia in academic and society research was
reviewed in \cite{wikiacad1,wikiacad2,wikiacad3,wikiacad4}.

In these WIKI networks we consider a confrontation and influence of
groups of opposite fixed opinions (spins) given by
nodes (articles) {\it capitalism} (blue color, $\sigma=-1$)
and {\it socialism}  (red color, $\sigma=1$)
and its extended case when each group is formed by
two by two nodes {\it capitalism, imperialism}
and {\it socialism, communism} (each language edition
determines these articles by corresponding transcription).
We also shortly consider interactions and influence of other two
opposite groups with fixed opinions/spins
given by articles {\it Christianity} and
{\it Islam}, and also
{\it Democratic Party (United States)}
vs {\it Republican Party (United States)}.
The description of data sets,
Monte Carlo procedure of spin interactions
and obtained results are presented in next Sections.

After the seminal work of Karl Marx in 1867 \cite{marx}
a great variety of  research investigations
appeared about the conflict between capitalism and socialism
being based on economics and sociological
science analysis (see e.g. \cite{capital1,capital2} and Refs. therein).
Here we use another purely mathematical and numerical analysis
of Wikipedia networks of 6 language editions
which allows us to determine the opinion preference
to socialism or capitalism in global for a whole edition
and also for specific articles of Wikipedia
such as world countries, historical political figures.
A clear meaning of each Wikipeadia article 
allows also to test the efficient and weak features of
our INOF approach. Since Wikipedia accumulates a huge
amount of human knowledge  \cite{wikiacad1,wikiacad2,wikiacad3,wikiacad4}
we think that the obtained results are of general public interest.

The article is composed as follows: Section 2 describes the Ising Network Opinion Formation (INOF) model,
data sets and numerical methods, Section 3 presents results for confrontation of
opinions for capitalism and socialism, Section 3 considers interactions between Christianity and Islam,
competition between US Democratic and Republican parties and studies in Section 4,
statistical features of the proposed INOF model are discussed in Section 5, 
discussion and conclusion are given in Section 6.\\

\textbf{ERRATUM added at July $31$, 2025}:
Due to a misprint all numerical results of this work
have been obtained with the following expression for
spin flip condition in Eq.(1):
$Z_i = \sum_{j \neq i} \sigma_{j} {A}_{ij} $
where $A_{ij}$ is the adjacency matrix of network
(see also related arXiv:2507.22254 [cs/SI])

\section{Model description and data sets}

We use Wikipedia networks of 2017 with their 6 language editions,
data sets are taken from \cite{wiki2017}.
Thus EN-wiki network has about $N \approx 5.4 \times 10^6$ nodes,
while others 5 networks have around $N \approx 1.3 (ES) - 2 (DE)$ million nodes;
exact number of nodes and links are given in \cite{wrwu2017}.

We characterize all network nodes by their
PageRank vector $P(i)$ probability  \cite{brin,meyer,rmp2015}
normalized to unity (${\sum_{i=1}}^N P(i) =1$); thus all nodes
get the PageRank index $K$ that orders
nodes by a monotonically decreasing probability $P(K)$
with highest probability at $K=1$ and smallest at $K=N$.
PageRank vector is the eigenvector of
the Google matrix $G$ \cite{brin,meyer,rmp2015}
with the highest eigenvalue $\lambda=1$:
$G P = \lambda P = P$ and $G= \alpha S +(1-\alpha)/N$.
Here $S_{ij}$ is the matrix of Markov transitions between nodes
constructed from adjacency matrix $A_{ij}$; thus
$S_{ij}=1/k_{ij}$ where $k_{ij}$ is a number of out-going links
from node $j$ to node $i$; for dangling nodes without out-going
links $S_{ij}=1/N$ and $\sum_i S_{ij} =1$. We use a standard value of the damping
factor $\alpha =0.85$ \cite{brin,meyer,rmp2015},
it regularizes the network connecting all isolated communities.

To determine the steady-state configuration of spins on a given network
we mainly follow an asynchronous Monte Carlo procedure describe in \cite{brics2024}
with an additional important modification.
The selected nodes (wiki-articles) have assigned fixed spin values
($\sigma_l=-1$ blue for {\it capitalism} and $\sigma_k=1$ red
for {\it socialism}, this is called option-1 (OP1);
or $\sigma_l=-1$ for {\it capitalism} and {\it imperialism}
and $\sigma_k=1$ for {\it socialism} and {\it communism},
this is called option-2 (OP2)). In a difference from \cite{brics2024}
all other nodes supposed to have a white color 
(or spin $\sigma=0$) at the initial stage of Monte Carlo process,
we call this a white option. Such a choice of initial state of all spins
corresponds to a situation when all other spins, those which are not
fixed,  have no definite opinion at initial stage.
Then by random we choose a spin $i$, which is not fixed,
and compute its influence score from in-going links $j$:
\begin{equation}
	Z_i = \sum_{j \neq i} \sigma_{j} {\tilde{S}}_{ij} .
	\label{eq1}
\end{equation}
where the sum is performed over all nodes $j$ pointing to $i$;
${\tilde{S}}_{ij} $ is the matrix $S_{ij}$ of Markov transitions
where the columns of dangling nodes have zero elements
(dangling nodes give no contribution to $Z_i$).
Also $\sigma_j=1$ if spin of $j$ node oriented up
or $\sigma_j=-1$ if it is oriented down or
$\sigma_j=0$ if node $j$ has no opinion (belongs to
initial set of white option). After the computation of value $Z_i$ a spin of node $i$
takes value $\sigma_i=1$ if $Z_i > 1$ or $\sigma_i=-1$ if $Z_i <0$
or stay unchanged if $Z_i=0$. Then such a random iteration
is done for another random node $i'$, without repetition for previously visited nodes.
We use a random shuffle  to perform this operation. 
Thus after $N$ such random iterations (fixed nodes remain fixed)
we make a full time step with time $\tau=1$ and then all procedure
is repeated going to $\tau=2,3,...$.
The process of convergence to a steady-state
is shown in Fig.~\ref{fig1}.
We find that at $\tau = 20$ 
the process is converged to a steady-state distribution of spins
with a fixed final fraction $f_r$ of red nodes with spins up
and a final $f_b$ of blue nodes with spins down.
There is a small fraction of nodes
that remains white at $\tau \geq 20$ that we attribute to
a presence of isolated communities \cite{rmp2015}.
However, a number of such nodes $N_{isol}$ is relatively small
(e.g. for OP2 we have
$N_{isol}/N \approx 
 0.135; 0.140; 0.117; 0.165; 0.109; 0.176$ for EN; DE; FR; 
 RU; IT; ES Wikipedia editions respectively).
We do not take into account these final white nodes from
isolated communities considering only red and blue nodes in the final steady-state
with a natural normalization of their fractions $f_r+f_b=1$.
We also characterize the final state by its polarization (or magnetization)
of spins given by
$\mu = f_r \sigma_{+} + f_b \sigma_{-} =2f_r - 1$.

However, we should note that in the Monte Carlo process one can choose various random ordering of
spin flip defined by the rule (\ref{eq1}) and
thus we obtain various random realisations
of pathway ordering of spins forming various random pathways
leading to a finial steady-state distribution. In fact we find that different
random pathways lead generally to different final configurations of spins
as it is shown in Fig.~\ref{fig1}. Due to that we perform an averaging over
$N_r=1000$ random pathway realisations (we call this 1000 pathways as a slot).
The histograms of fractions of red nodes obtained from $N_r=1000$
realisations are shown in Fig.~\ref{fig2} and Fig.~\ref{fig3}
for Wikipedia editions and options OP1 and OP2 respectively.
By making average over these random realisations
we obtain the steady-state values of $f_r(i), f_b(i), \mu_i$
for each node (spin) $i$. By definition $-1 \leq \mu_i \leq 1$.
 Thus after averaging over all $N_r$ realisations each node $i$
is characterized by its average values $f_r(i)$ (we will speak mainly about
fraction of red nodes),  $\mu_i$ and deviation from global polarization
$\Delta \mu_i = \mu_i - \mu_0$. 
After averaging over all nodes we obtain global
network values of red and blue node $f_r, f_b$ and global network polarization
$\mu_0 = <\mu_i>$. We checked that the probability distributions of Figs.~\ref{fig2},\ref{fig3}
remain unchanged if we increase the time $\tau$ from $\tau=20$ to $\tau=40$.
Thus all the realists are take from the steady-state at $\tau=20$.
The results with increased number of realisations, up to $N_r = 10^5$
are discussed in Section 6.

We note that in the relation for $Z_i$ in (\ref{eq1}) we use only
matrix elements $S_{ij}$ without dangling nodes. The reason for this choice is
due to a fact that matrix elements (or their part)
that are the same for all nodes in a column or in the whole matrix
(as in $G$ matrix with $(1-\alpha)/N$ term) act similar to a certain
external magnetic (polarization) field
that gives a contribution proportional to a difference of red and blue node fractions
while we aim to analyze interactions between node spins without external fields.
In the sum of (\ref{eq1}) we include only contributions
of in-going links given by $S_{ij}$ since in Wikipedia networks
in-going links are more robust while out-going links are characterized
by significant fluctuations \cite{rmp2015}. In this sense this is different from
trade networks where both in-going and out-going links are important
corresponding to import and export \cite{brics2024}. In our case (\ref{eq1})
all $S_{ij}$ are positive or zero that corresponds to
some kind ferromagnetic interactions between spins.
However, a presence of fixed spins of opposite orientations
makes possible to have big configurations of spins oriented
up or down. It is useful to note that a similar type of
relation (\ref{eq1}) is used in models of associative memory
however there the elements $S_{ij}$ take random values $\pm 1$
corresponding to some kind of anti-ferromagnetic
interactions \cite{memory1,memory2}; but fixed spins
and white option for nodes are not considered there.

We call the above approach of opinion formation on directed networks as
Ising Network Opinion Formation (INOF) model.

\section{Results for capitalism vs. socialism}

\begin{figure}[H]
	\begin{center}
		\includegraphics[width=0.7\columnwidth]{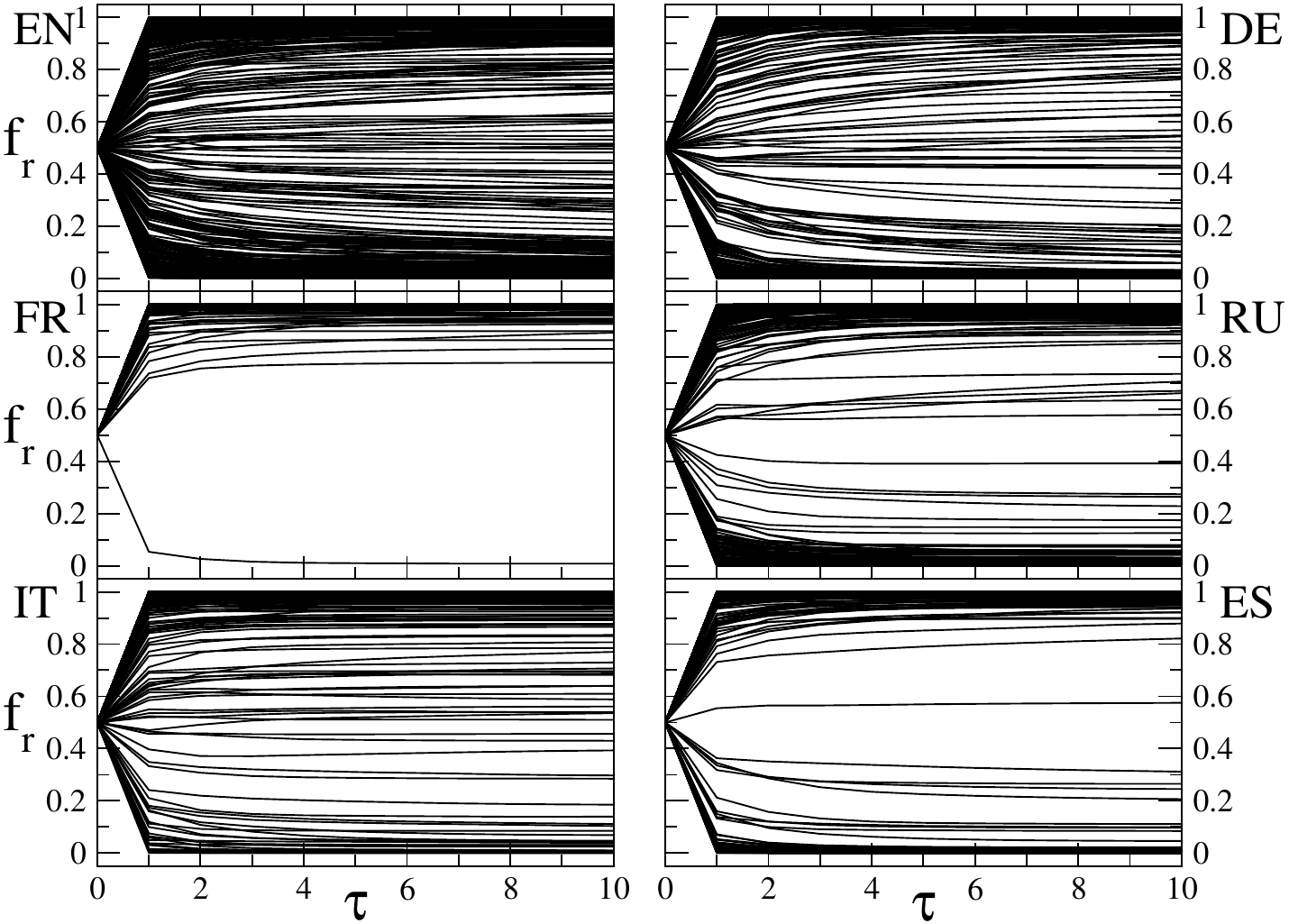}
	\end{center}
	\vglue -0.3cm
	\caption{\label{fig1}Evolution of the fraction of red nodes 
  $f_r$ for 500 random pathway realisations. An initial 
  condition has  fixed 2 red nodes ({\it socialism,communism}) and 2 blue nodes 
  ({\it capitalism,imperialism}), they remain fixed during the Monte Carlo evolution
  with the relation (\ref{eq1}), all other nodes are initially white.
      Each panel corresponds to one of the 
  six  language editions of Wikipedia: EN (English), DE (German),
  FR (French), RU (Russian), IT (Italian), and ES (Spanish). 
  The x-axis represents time $\tau$ , where each unit of $\tau$ 
  indicates one complete update of all nodes/spins following the opinion model
  based on relation (\ref{eq1}).
	}
\end{figure}

In Fig.~\ref{fig1} we show a convergence
with time $\tau$ to a steady-state values of
fraction of red nodes $f_r$  
($f_r=1$ corresponds to preference to {\it socialism,  communism}
and spin up orientation, $f_r=0$ to {\it capitalism, imperialism}).
Data are given for 500
random pathway realisations for each of  6 language editions of Wikipedia of
year 2017. The results show that the steady-state values are reached
at $\tau=10$, to be  completely sure
we show in further the steady-state values taken at $\tau=20$.

The realisations shown in Fig.~\ref{fig1} indicate that
the red-preference to {\it socialism, communism}
is different for each of 6 Wikipedia editions:
thus EN case has a comparable number of cases with final $f_r$ close to unity or zero
while FR case has mainly $f_r$ close to unity with $f_b=1-f_r $ close to zero.

\subsection{Statistical properties of opinion polarization}

\begin{figure}[H]
	\begin{center}
		\includegraphics[width=0.7\columnwidth]{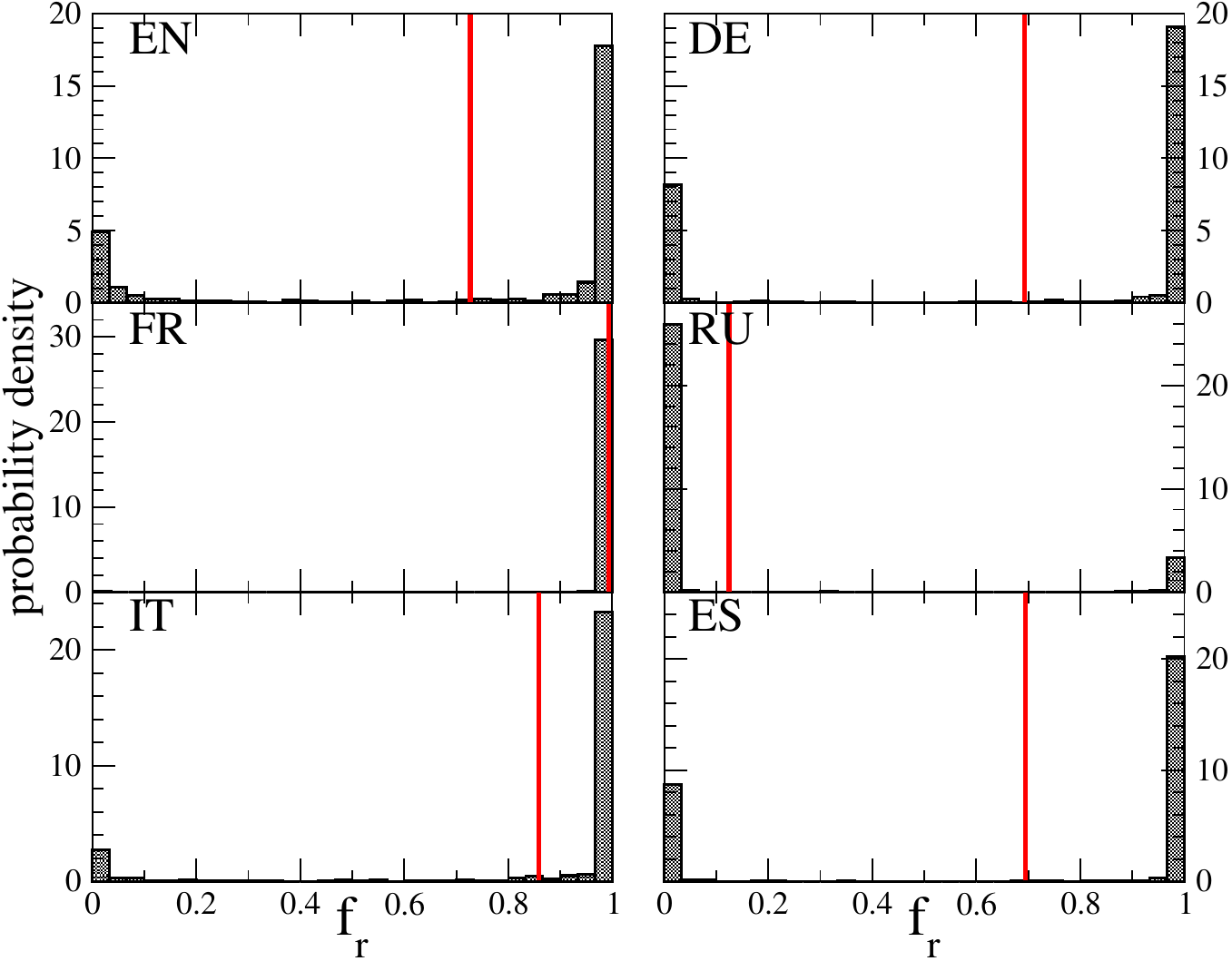}
	\end{center}
	\vglue -0.3cm
	\caption{\label{fig2} Probability density $p$ of the fraction of 
  red nodes ($f_r$) for $1000$ realisations after $\tau=20$. Each panel for OP1
  corresponds to one of the six different languages of Wikipedia for the initial 
  condition with fixed one red node ({\it socialism}) and one blue node 
  ({\it capitalism}). 
  Red vertical lines mark the mean value of $f_r$ with average global  polarization 
  $\mu_0=2f_r-1$.
  The values of mean polarization $\mu_0$ are: 
  $0.455$ for EN, $0.385$ for DE,  $0.389$ for ES,
  $0.986$ for FR,  $0.717$ for IT and $-0.752$ for RU. 
  The histogram is built with cell size $\Delta f_r = 1/30$ and normalized to 
  1 ($\sum_{f_r} p \Delta f_r = 1$).
	}
\end{figure}

\begin{figure}[H]
	\begin{center}
		\includegraphics[width=0.7\columnwidth]{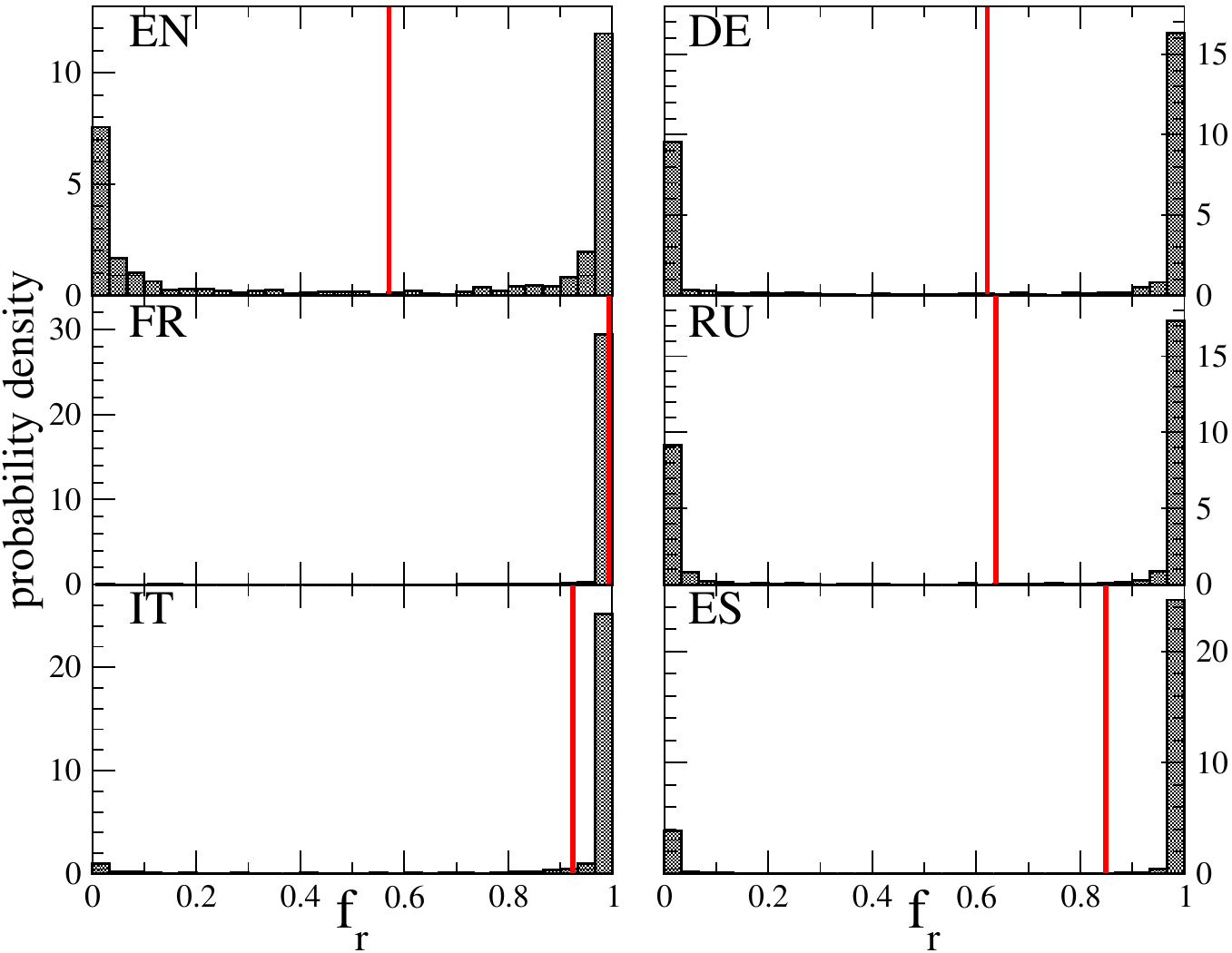}
	\end{center}
	\vglue -0.3cm
	\caption{\label{fig3} Same as in Fig.~\ref{fig3}, each panel for OP2 
  corresponds to one of the six different languages of Wikipedia for the initial 
  condition with fixed two red nodes ({\it socialism, communism}) and two blue nodes 
  ({\it capitalism, imperialism}). The values of mean polarization $\mu_0=2f_r-1$ are: 
  $0.141$ for EN, $0.243$ for DE,  $0.697$ for ES,
  $0.986$ for FR,  $0.849$ for IT and $0.276$ for RU. 
  The histogram is built with cell size $\Delta f_r = 1/30$ and normalized to 
  1 ($\sum_{f_r} p \Delta f_r = 1$).
	}
\end{figure}

To understand the statistical properties of various configurations
we present in Fig.~\ref{fig2} and Fig.~\ref{fig3} the histograms showing the
frequency of appearance of final steady-state values of $f_r$
obtained from one slot of $N_r=1000$ realisations. At first we stress
that almost all final configurations have all nodes being red for FR-wiki
with average $<f_r>=1$ ($\mu_0 \approx 1$).
Vary similar situation takes place for IT and ES editions
with average red fractions given in Fig.~\ref{fig2}, Fig.~\ref{fig3} captions.
The situation is more balanced for EN, DE, RU editions.
Thus the results of Fig.~\ref{fig3}
show that EN, DE, RU editions have certain preference to {\it capitalism, imperialism}
even if their preference to {\it socialism, communism} is stronger.
In contrast the Wikipedia editions  ES, FR, IT have their preference almost
completely for {\it socialism, communism}. We should note that the steady-state
almost for each realisation is composed only from all red or all blue nodes
(only for EN there is relatively small number of final configurations
which have both red and blue nodes; a number of such mixed configurations
i very small for other editions). Such a situation is very different
from results obtained for the world trade networks \cite{brics2024}
where stead-state configurations had high fractions of red and blue nodes.
We attribute this to a different internal structures of
Wikipedia and trade networks.

It is interesting to compare the results for the case OP1 of Fig.~\ref{fig2}
with fixed {\it socialism} (red) and {\it capitalism} (blue)
with the case of OP2 in Fig.~\ref{fig3} when we have fixed {\it socialism, communism} (red)
and {\it capitalism, imperialism} (blue). For 5 editions OP1 case has significantly
higher red fractions $f_r$ comparing to the OP2 case. Thus addition of
fixed red node {\it communism} and blue one {\it imperialism}
plays against red opinion. However, the situation is drastically different
for RU-wiki: for OP1 it has very strong preference for
{\it capitalism} while for OP2 case it has stronger
preference for {\it socialism, communism}.
We attribute this result to the fact that from 1917 till 1992 Russia (or USSR)
was ruled by the Communist party which had an official aim to built communism.
We note that  in EN edition the article {\it Russia} has
in-going link from
{\it communism} and {\it imperialism},
but not from {\it socialism} and {\it capitalism}.
In RU edition it has no in-going from
{\it socialism, communism. capitalism, imperialism}
that probably makes it more influenced by other longer pathways from fixed nodes.
Thus probably the imperial period of Russian history,
being significantly longer comparing to Soviet period,
produces a certain trend to blue fraction
(see discussion for China below).

In fact Russian Wikipedia was established after disappearance of USSR
and thus it has not so strong stress on political formations.
Also  the period followed after USSR in 1991-2000 is known
in Russia as a period of ``wild capitalism''
that is probably at the origin of strong preference to {\it capitalism} for OP1
in RU edition.

\begin{figure}[H]
	\begin{center}
		\includegraphics[width=0.6\columnwidth]{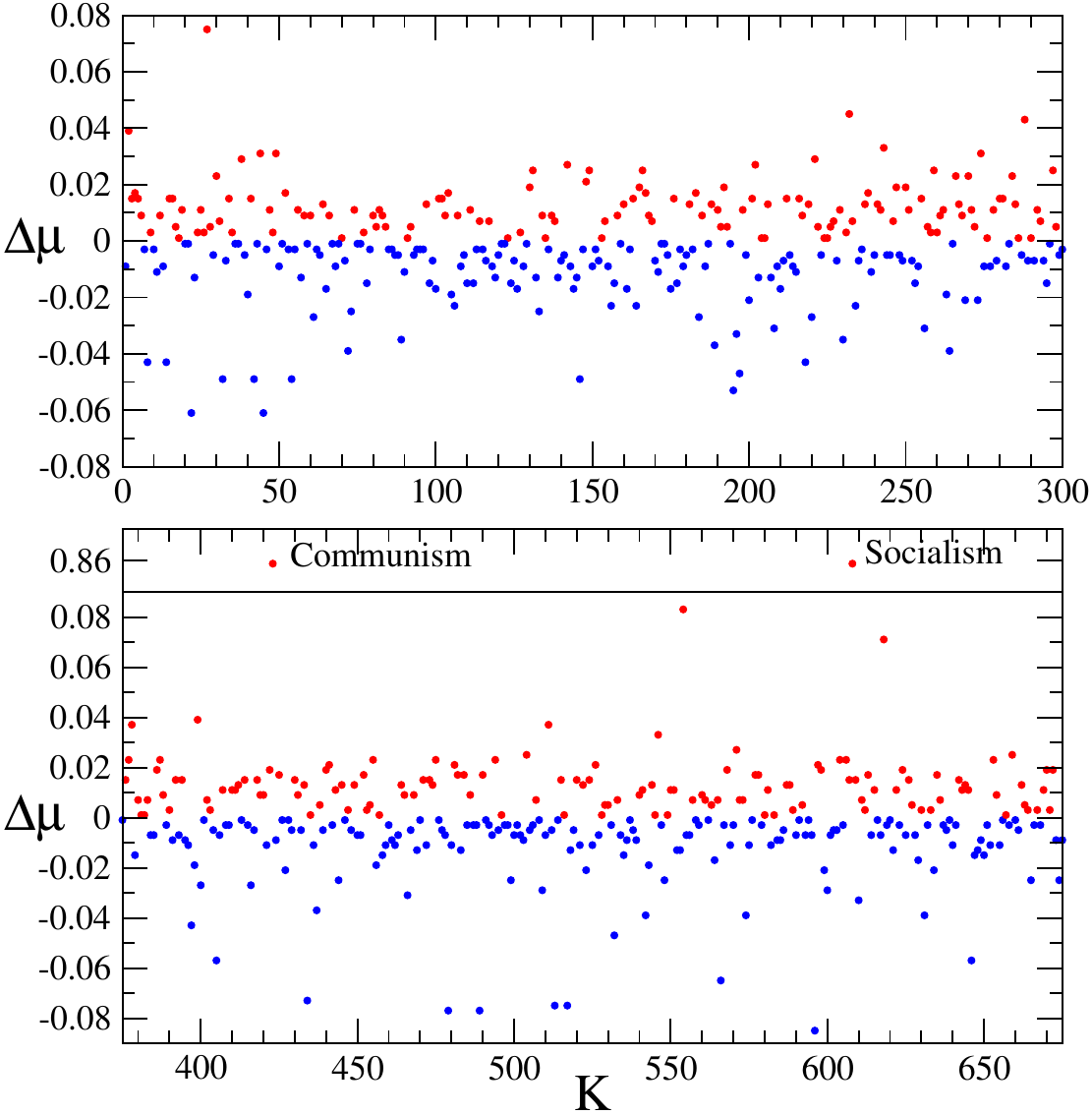}
	\end{center}
	\vglue -0.3cm
	\caption{\label{fig4} 
          Average polarization values for the top $K$ nodes ($\Delta \mu = \mu - \mu_0$)
          for EN edition and OP2 case; here $\mu_0=0.141$.
  Positive and negative $\Delta \mu$ are represented by red and blue circles, 
  respectively. The top panel shows the case for the top 300 PageRank ranks ($K$), 
  while the bottom panel displays ranks from 375 to 675, where the 
  "Communism" and "Socialism" nodes appear in English language. 
  The average is computed over 1000 iterations after $\tau=20$.
	}
\end{figure}

By averaging spin $\sigma_i$ of each article/node $i$ over $N_r=1000$ reslisations
we obtain average polarization $\mu_i$  of node $i$.
Ordering all nodes $i$ by the PageRank index
$K$ we obtain dependence of polariization $\mu(K)$ on $K$. For EN Wikipedia
this dependence $\Delta \mu(K) = \mu(K) - \mu_0$ is shown in Fig.~\ref{fig4} for OP2
for top PageRank indexes $1 \leq K \leq 300$ (top panel)
and for the range $375 \leq K \leq 675$ which contains articles {\it socialism, communism}
located at  $K=608, 423$; {\it capitalism, imperialism} are located at $K=831,3904$.
Comparing to the average global polarization $\mu_0$ each article has its own
$\Delta \mu$ deviation shown in Fig.~\ref{fig4}. Typically we have these deviations in the range
$-0.1 < \Delta \mu <0.1$ with some exceptional deviations (of course fixed 4 nodes
have higher $\Delta \mu $ absolute values).
We discuss these deviations for specific articles (nodes) below in next subsection.

It can be possible to expect that an average opinion polarization $\mu_0$ for a give edition
is  related with PageRank probabilities $P_r$ and $P_b$ of fixed red and blue nodes
(with rescaled sum equal to unity $P_r + P_b =1$) with $\mu_0 \approx 2P_r -1$.
For our 6 editions and OP2 case the values of $P_r$ are located in a relatively narrow range $0.66 < P_r <0.73$
while the values of $\mu_0$ are dispersed in the range  $0.15 < \mu_0 <1$
without any clear correlation with $P_r$ values.
For OP1 case we have the range $0.5 < P_r < 0.7$
and $-0.24 < \mu_0 < 1$ again without any clear correlation between $P_r$ and $\mu_0$ values.
Thus we conclude that there is no
correlation between $\mu_0$ and $P_r$.

\begin{figure}[H]
	\begin{center}
		\includegraphics[width=0.6\columnwidth]{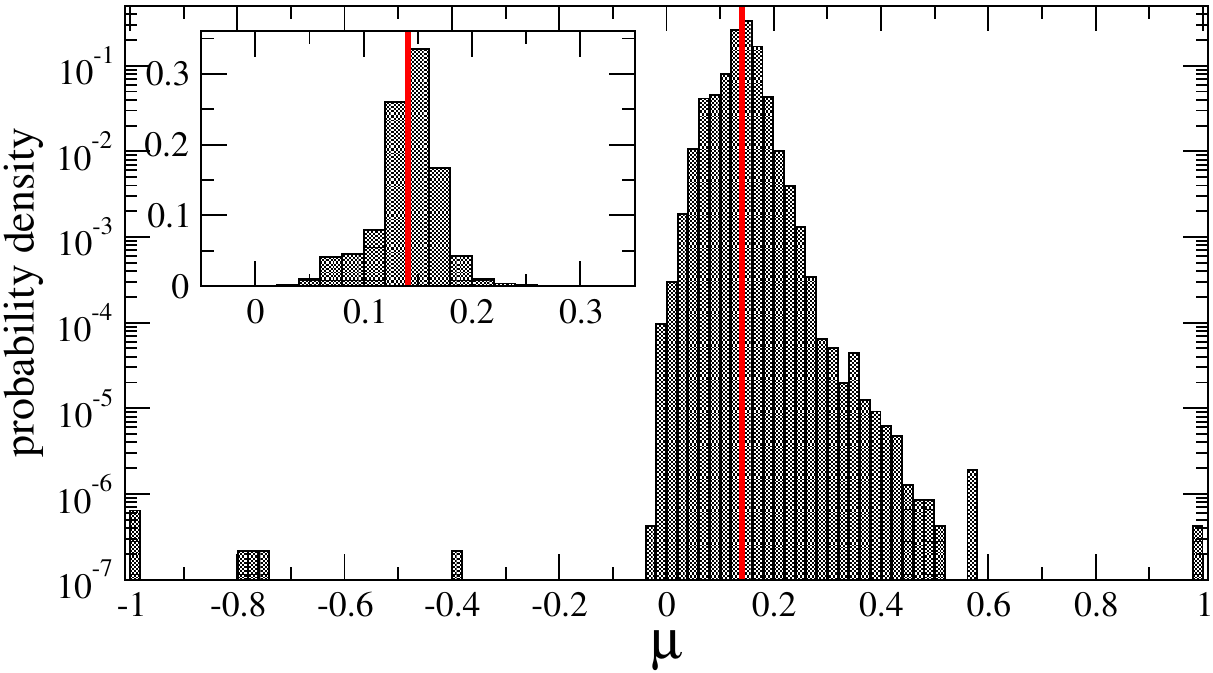}
	\end{center}
	\vglue -0.3cm
	\caption{\label{fig5} 
          Probability density of the average polarization value $\mu$
          for the English edition for OP2 case; here $\mu_0=0.141$. 
  The main panel displays the probability density on a logarithmic scale, while the inset 
  panel shows it on a linear scale. The average is computed over 1000 realisations and $\tau=20$.
	}
\end{figure}

After averaging over $N_r$ random pathway realisations we obtain
opinion polarization $\mu$ for all  Wikipedia articles.
The distribution histogram or probability density $p$ for these $N$
polarization $\mu$ values  is shown in Fig.~\ref{fig5} for EN edition and OP2.
The main density if concentrated in the range
$-0.05 < \mu < 0.55$ centered around the global average polarization $\mu_0 = 0.141$. 
We discuss specific articles with extreme positive or negative $\Delta \mu$
values below in next subsection.

\begin{figure}[H]
	\begin{center}
		\includegraphics[width=0.6\columnwidth]{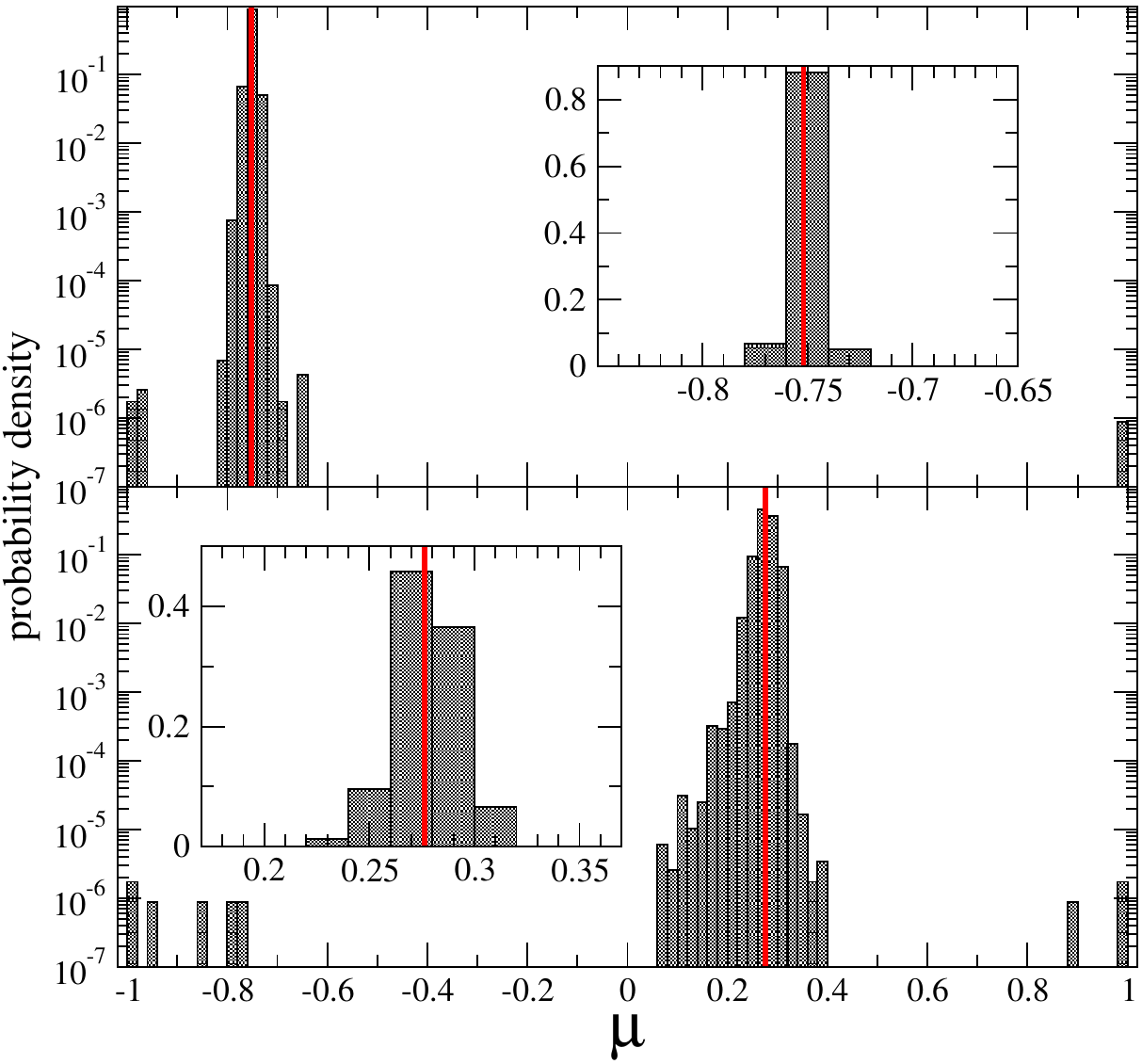}
	\end{center}
	\vglue -0.3cm
	\caption{\label{fig6} 
          Probability density of the average magnetization value $\mu$ for the Russian
          edition for OP1 (top panel) and OP2 (bottom panel) OP2 case;
          here $\mu_0=-0.752; 0.276$ respectively.
  The main panels display the probability density on a logarithmic scale.
  Top panels represent the initial  condition with one fixed red node ({\it socialism}) 
  and one fixed blue node 
  ({\it capitalism}) on top panels, while bottom panels show 
  the initial condition with 2 fixed red nodes ({\it socialism,communism}) and with 2 fixed blue nodes 
  ({\it capitalism,imperialism}),
  The  corresponding inset 
  panels show the same probability density but on a linear scale. 
  The average is computed over 1000 realisations and  $\tau=20$.
	}
\end{figure}

Similar to the case of EN edition with Fig.~\ref{fig5} we show such histograms
for RU edition for both options OP1 and OP2 in Fig.~\ref{fig6}. These histograms
clearly demonstrate the drastic difference between OP1 and OP2 cases
which we attribute to the ruling Communist party of Russia as we pointed above.

\begin{figure}[H]
	\begin{center}
		\includegraphics[width=0.55\columnwidth]{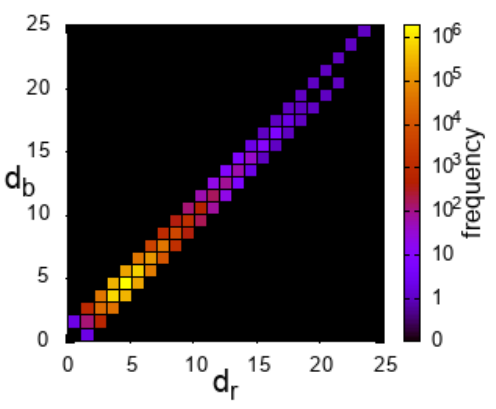}
	\end{center}
	\vglue -0.3cm
	\caption{\label{fig7} 
  Distribution of English Wikipedia articles based on their distance to red nodes 
  {\it socialism, communism} and blue nodes {\it capitalism, imperialism}. 
  Color represents the frequency/number of articles as a function of 
  $(d_r,d_b)$. 
	}
\end{figure}

It is natural to expect that the  Erd\"os number \cite{dorogovtsev10} or
Erd\"os link distance (number of links) from red and blue groups of fixed nodes
(we discuss OP2 case for EN edition) should significantly influence
the opinion formation on the Wikipedia network. To analyze this feature we
show in Fig.~\ref{fig7} a number of network nodes $N_d$ (or frequency)
located on distances $d_r$ from two fixed red nodes {\it socialism, communism}
and  $d_b$ from two fixed blue nodes {\it capitalism, imperialism}.
The number of such nodes $N_r$ grows exponentially with
distance up to values $d_r \approx d_b \approx 6$ where there are
up to million nodes $(N_d \approx 10^6)$; for larger $d_r, d_b$
values $N_d$ decreases since due to the small world effect \cite{dorogovtsev10}
majority of network nodes can be reached in $d_r, d_b \leq 6$ links
(degree of separation). The interesting feature of Fig.~\ref{fig7}
is that all $N_d$ nodes are located on three diagonals
with $d_b=d_r, d_b=d_r \pm 1$. We argue that both groups of fixed red and blue nodes
describe the human society and thus there are close relations (small number
of links) between these two groups. Indeed, for EN edition the Erd\"os distance
between these two groups is 1 . We find the same three diagonal structure
as in Fig.~\ref{fig7} for other 5 editions. 

\begin{figure}[H]
	\begin{center}
		\includegraphics[width=0.7\columnwidth]{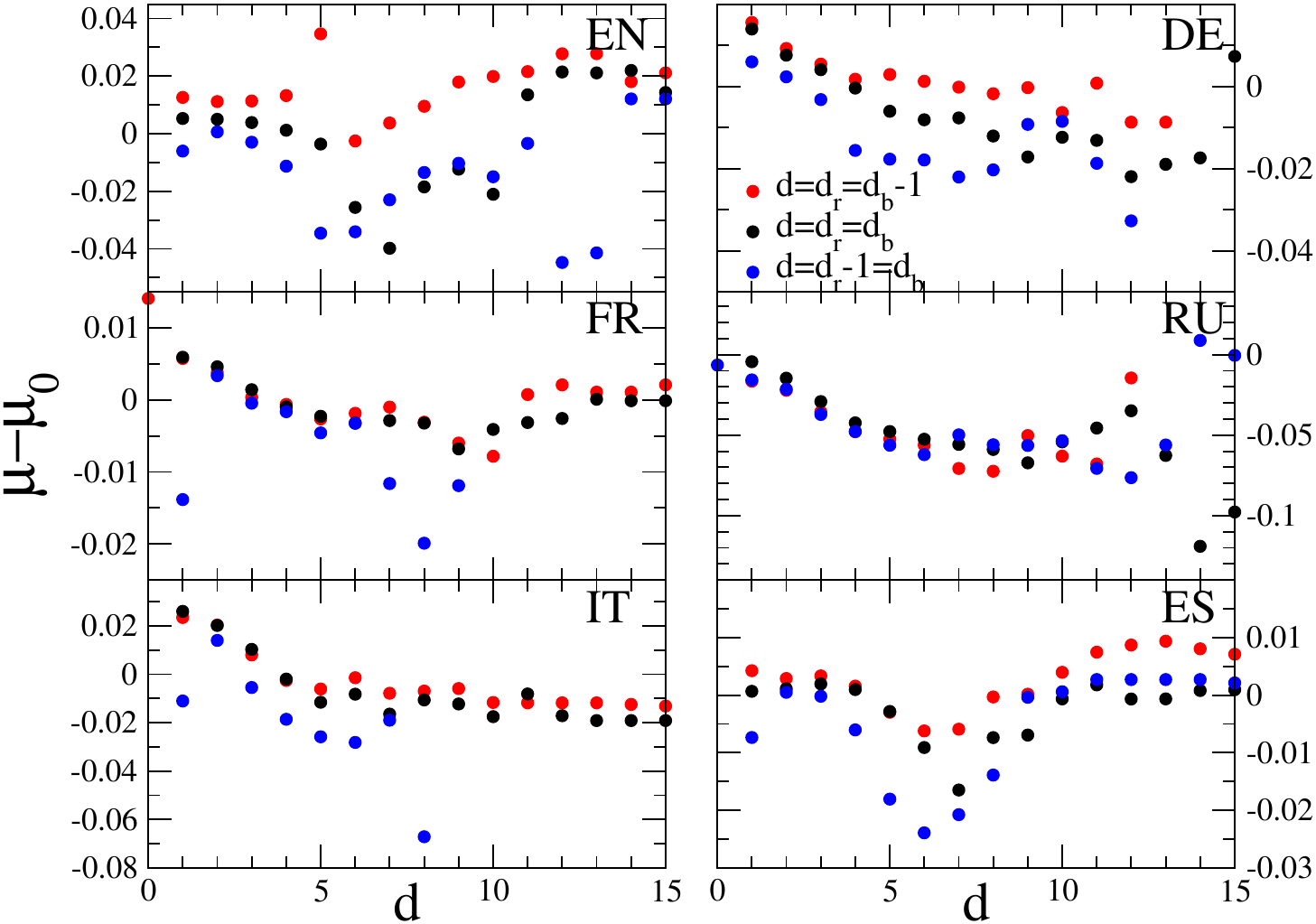}
	\end{center}
	\vglue -0.3cm
	\caption{\label{fig8} 
  Average polarization $\Delta\mu=\mu-\mu_0$ as a function of Erd\"os distance $d$ for OP2 case. 
  Each panel corresponds to one of the six different languages of Wikipedia: 
  EN (English), DE (German),  ES (Spanish), FR (French),  IT (Italian) and RU (Russian). 
  Red circles represent nodes that are one step closer to red nodes than to blue nodes
  ($d=d_r=d_b-1$),
   black circles represent nodes that are equidistant from red and blue nodes ($d=d_r=d_b$), 
   and blue circles represent nodes that are one step closer to blue nodes than to red nodes 
  ($d=d_r-1=d_b$). The average values $\mu_0$ are given in caption of Fig.~\ref{fig3}.
	}
\end{figure}

In Fig.~\ref{fig8} we show the average polarization $\Delta \mu = \mu - \mu_0$
for each cell located at
Erd\"os distances $(d_r, d_b)$ along the three diagonals for all 6 editions.
The results show that on average for moderate $d$ distance ($d \leq 6$)
we have $\Delta \mu$ being larger when the distance to the red group
is shorter than to blue  group. This is also well visible for ED,DE, ES, IT editions
while for FR and RU editions this difference is less pronounced.

\subsection{Opinion polarization of specific articles}

We discuss now the opinion polarization of specific articles
concentrating mainly on EN edition of Wikipedia.
Thus in Table~\ref{tab1} we present top 20 PageRank articles
with their polarization opinions $\Delta \mu = \mu - \mu_0$
for two options OP1, OP2 of fixed nodes. In fact the average global
opinion polarization $\mu_0$ of the whole network
gives the average polarization background (see e.g.  Fig.~\ref{fig4})
and thus it is more informative to present the deviations
from this background given by $\Delta \mu$.

\begin{table}[h!]
  \centering
  \begin{tabular}{|c|l|c|c|c|}
  \hline
  K & Title & $\Delta\mu$ (OP1) & $\Delta\mu$ (OP2) & $\Delta\mu(L)$ (OP2) \\
  \hline
  1	&	United States	&	-0.0086	&	-0.009	&	0.002	\\
  2	&	Association football	&	0.013	&	0.039	&	0.034	\\
  3	&	World War II	&	0.021	&	0.015	&	0.013	\\
  4	&	France	&	0.023	&	0.017	&	0.015	\\
  5	&	Germany	&	0.019	&	0.015	&	0.016	\\
  6	&	United Kingdom	&	0.025	&	0.009	&	0.014	\\
  7	&	Iran	&	0.027	&	-0.003	&	0.002	\\
  8	&	India	&	-0.0066	&	-0.043	&	-0.034	\\
  9	&	Canada	&	0.0014	&	0.003	&	0.006	\\
  10	&	Australia	&	0.0094	&	-0.003	&	0.002	\\
  11	&	China	&	0.021	&	-0.011	&	-0.001	\\
  12	&	Italy	&	0.025	&	0.009	&	0.013	\\
  13	&	Japan	&	0.017	&	-0.009	&	-0.003	\\
  14	&	Moth	&	-0.0046	&	-0.043	&	-0.039	\\
  15	&	England	&	0.023	&	0.015	&	0.016	\\
  16	&	World War I	&	0.023	&	0.015	&	0.013	\\
  17	&	Russia	&	0.025	&	0.005	&	0.014	\\
  18	&	New York City	&	0.0014	&	0.001	&	0.009	\\
  19	&	London	&	0.017	&	0.011	&	0.014	\\
  20	&	Latin	&	0.025	&	-0.001	&	0.007	\\
  \hline
  \end{tabular}
  \caption{Top 20 PageRank index $K$ articles of English Wikipedia and $\Delta \mu$ for fixed one
  red node ({\it socialism}) and one fixed blue node 
  ({\it capitalism}) (OP1); and for fixed two  red nodes 
  ({\it socialism,communism}) and two blue nodes 
  ({\it capitalism,imperialism}) (OP2) with a slot of $10^3$ realisations, 
  and $\Delta \mu(L)$ for OP2 long run of $10^5$ realisations . Here $\Delta \mu = \mu - \mu_0$
  where $\mu$ is polarization of given article and $\mu_0 = 0.455; 0.243, 0.146; $ is the average
  global polarization of EN Wikipedia 2017 for OP1 and OP2 respectively.
  The values of $\Delta \mu(L)$ are discussed in Section 6. }
  \label{tab1}
  \end{table}

There are only 3 articles with negative $\Delta \mu$ values for PageRank top 20 articles
in Table~\ref{tab1} for OP1
while for OP2 case there 8 such cases. This approximately
corresponds to a significantly higher peak at $f_r=0$ for OP2 in
Fig.~\ref{fig3} compared to OP1 case in Fig.~\ref{fig2} for EN edition.
The main part of this top 20 PageRank list in Table~\ref{tab1}
is composed with world countries.
Among other type of articles we note that Association football, World War I, II
have positive $\Delta \mu$ values comparing to the global positive $\mu_0$ value
of EN edition network, while $Moth$ has a negative $\Delta \mu$.
Due to many in-going links to top PageRank articles it is difficult to
identify the origins of such polarization opinion for these articles.

For the case of world countries we present an additional Table~\ref{tab2}
for OP2 case showing top 20 countries from the PageRank global list.
Here we find that all European countries from this list of 20
have positive $\Delta \mu$ (including Russia). In contrast to that
other countries outside of this area have negative $\Delta \mu$
(except Canada and Mexico). We suppose that positive $\Delta \mu$
for European countries is related to the fact that socialism concept
was developed in these countries. The reasons of negative $\Delta \mu$
values for  Japan and Brazil requires mode deep analysis
of network link structure.
For China and India we explain negative $\Delta \mu$  
in the following way.

Thus for China we note from Table~\ref{tab1} that $\Delta \mu$ is positive
for OP1 case and negative for OP2 case that seems to be somewhat surprising
in view of strong influence of Communist party in China.
We argue that this appears due to the fact that
the word {\it imperialism} becomes
in Wikipedia linked with words {\it imperium, empire, emperor, imperator}.
Indeed, the articles  in Fig.~\ref{fig5}
at an extreme negative
values with $\mu = -0.022$
is a serviteur Li Yong (chancellor) of China Emperor Xianzong in 8  century AD.
Thousands of years China was a powerful {\it empire}
that, in our opinion, shifts China to negative $\Delta \mu$ value.
Also China is directly pointed by articles {\it imperialism, capitalism}
but also by {\it socialism, communism}.
We not that it happens rather often that an article
is pointed by fixed articles of opposite opinions.

For India we find that this article is directly pointed
by {\it imperialism} but not by
{\it capitalism, socialism, communism}.
We attribute such a difference in links to imperialism and colonisation
of UK in respect to India. We also note that for OP1, without imperialism,
the value of $\Delta \mu =-0.0066$ is still negative but it is significantly
smaller in absolute value comparing to OP2 case with imperialism and $\Delta \mu =-0.043$.
Thus we think this is the reason of highly negative $\Delta \mu$ value for India.

\begin{figure}[H]
	\begin{center}
		\includegraphics[width=0.7\columnwidth]{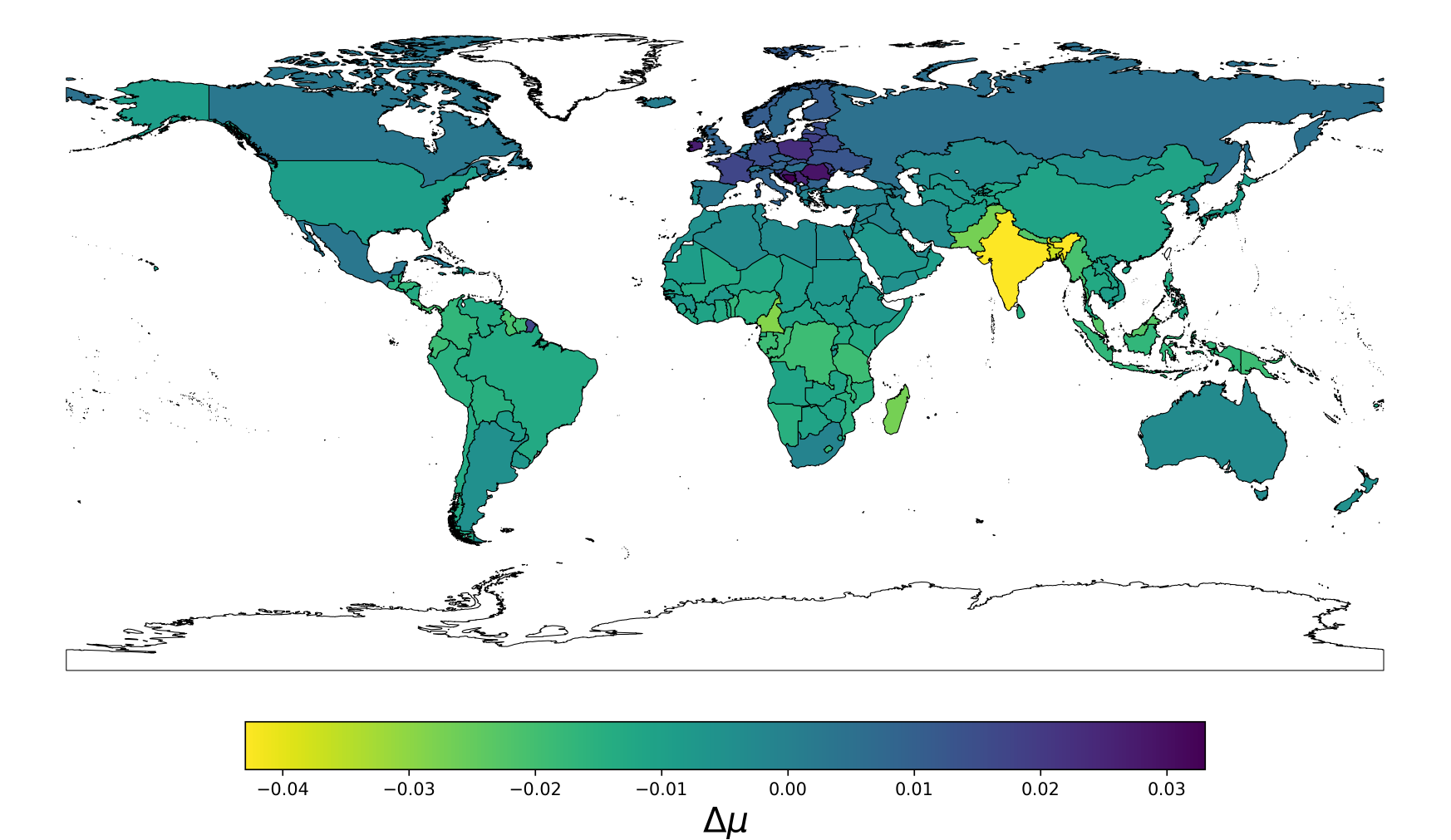}
	\end{center}
	\vglue -0.3cm
	\caption{\label{fig9} 
          Geographical distribution of opinion polarization
          to {\it socialism, communism} ($\Delta \mu >0$) or
          ({\it capitalism, imperialism}) ($\Delta \mu <0$) expressed by $\Delta \mu$
          for English Wikipedia.
          Color legend shows the scale for $\Delta \mu$.
	}
\end{figure}

  \begin{table}[h!]
    \centering
    \begin{tabular}{|c|c|l|c|c|}
    \hline
  $K_c$	&	K	&	Country	&	$\Delta\mu$ (OP2)	&	$\Delta\mu(L)$ (OP2) \\
  \hline
  1	&	1	&	United States	&	-0.009	&	0.002	\\
  2	&	4	&	France	&	0.017	&	0.015	\\
  3	&	5	&	Germany	&	0.015	&	0.016	\\
  4	&	6	&	United Kingdom	&	0.009	&	0.014	\\
  5	&	7	&	Iran	&	-0.003	&	0.002	\\
  6	&	8	&	India	&	-0.043	&	-0.034	\\
  7	&	9	&	Canada	&	0.003	&	0.006	\\
  8	&	10	&	Australia	&	-0.003	&	0.002	\\
  9	&	11	&	China	&	-0.011	&	-0.001	\\
  10	&	12	&	Italy	&	0.009	&	0.013	\\
  11	&	13	&	Japan	&	-0.009	&	-0.003	\\
  12	&	17	&	Russia	&	0.005	&	0.014	\\
  13	&	23	&	Brazil	&	-0.013	&	0.003	\\
  14	&	24	&	Spain	&	0.003	&	0.013	\\
  15	&	26	&	Netherlands	&	0.003	&	0.013	\\
  16	&	30	&	Poland	&	0.023	&	0.021	\\
  17	&	31	&	Sweden	&	0.007	&	0.015	\\
  18	&	35	&	Mexico	&	0.003	&	0.001	\\
  19	&	36	&	Turkey	&	-0.001	&	0.006	\\
  20	&	38	&	Romania	&	0.029	&	0.022	\\
  \hline
  \end{tabular}
  \caption{Top 20 Countries ($K_c$) given by PageRank index ($K$) in English Wikipedia
   and $\Delta \mu = \mu - \mu_0$ for the case of fixed two red nodes 
  ({\it socialism,communism}) and two blue nodes 
  ({\it capitalism,imperialism}) (OP2) for $10^3$ realisations and $10^5$ realisations: 
   $\mu_0=0.243$ and $\mu_0(L)=0.146$.
  The values of $\Delta \mu(L)$ are discussed in Section 6.}
  \label{tab2}
  \end{table}

  The opinion preference of all world countries
  to {\it capitalism, imperialism} or {\it socialism, communism},
  expressed by $\Delta \mu$ of countries, is shown in the world map of Fig.~\ref{fig9}.
  Positive opinions for {\it socialism, communism} with $\Delta \mu > 0$
  are located mainly in Europe, Russia, Canada and Mexico. The highest
  positive $\Delta \mu$ values are for
  Bosnia and Herzegovina ($0.033$), Romania ($0.029$), 
   Ireland ($0.027$), Poland ($0.025$), Croacia ($0.025$) and Serbia ($0.025$).
  The most strong opinions for {\it capitalism, imperialism}h with 
  lowest negative $\Delta \mu$
  are for India ($-0.043$), Bangladesh ($-0.037$), 
  Cameroon ($-0.029$), Pakistan ($-0.027$) and Madagascar ($-0.027$). 

  We also consider preference opinion for a group of historical figures, mainly politicians,
  presented in Table~\ref{tab3} for EN edition.
  They are composed of two groups of 12 figures in each group; the left group
  in Table~\ref{tab3} is composed by socialist-communist  active leaders
  and politicians (left column) and  the right group lists the
  capitalist political leaders (right column). Indeed, the obtained results
  show that for all socialist leaders in the left column  we obtain $\Delta \mu >0$
  (enhanced preference for socialism) (left column in Table~\ref{tab3}, except
  Mao Zedong for whom $\Delta \mu$ is only slightly negative being close to zero).
  In opposite, to capitalist leaders we obtain negative $\Delta \mu$
  corresponding to enhanced preference for capitalism (see right column of Table~\ref{tab3}
  with exception for Winston Churchill and Charles de Gaulle. We attribute
  these 3 exceptions to the fact
  all of them are influenced by their countries: Mao Zedong is linked with
  China having $\Delta \mu <0$;  Winston Churchill and Charles de Gaulle
  are linked with UK and France with $\Delta \mu >0$. We also note
  that other leaders in Table~\ref{tab3} are linked with Russia with $\Delta \mu >0$
  (left column; except Karl Marx) and with USA with $\Delta \mu < 0$.
  Thus the proposed method correctly determines preference opinion
  of leaders of socialism and capitalism.

  \begin{table}[h!]
    \centering
    \begin{tabular}{|l|c|c|l|c|c|}
      \hline
      Name	&	$\Delta\mu$	& $\Delta\mu(L)$ &	Name	&	$\Delta\mu$	& $\Delta\mu(L)$ \\
    \hline
    Karl Marx	&	0.007	&	0.018	&	Winston Churchill	&	0.017	&	0.015	\\
    Vladimir Lenin	&	0.017	&	0.021	&	Franklin D. Roosevelt	&	-0.003	&	0.006	\\
    Leon Trotsky	&	0.023	&	0.023	&	John F. Kennedy	&	-0.001	&	0.007	\\
    Joseph Stalin	&	0.019	&	0.020	&	Richard Nixon	&	-0.007	&	0.006	\\
    Nikita Khrushchev	&	0.015	&	0.019	&	Jimmy Carter	&	-0.005	&	0.006	\\
    Leonid Brezhnev	&	0.017	&	0.020	&	Ronald Reagan	&	-0.007	&	0.006	\\
    Yuri Andropov	&	0.017	&	0.021	&	George H. W. Bush	&	-0.007	&	0.005	\\
    Mikhail Gorbachev	&	0.013	&	0.019	&	Bill Clinton	&	-0.007	&	0.006	\\
    Boris Yeltsin	&	0.017	&	0.022	&	George W. Bush	&	-0.007	&	0.006	\\
    Vladimir Putin	&	0.013	&	0.020	&	Barack Obama	&	-0.007	&	0.006	\\
    Mao Zedong	&	-0.003	&	0.006	&	Donald Trump	&	-0.007	&	0.006	\\
    Xi Jinping	&	0.003	&	0.005	&	Charles de Gaulle	&	0.013	&	0.018	\\
\hline
\end{tabular}
    \caption{Historical figures of English Wikipedia, mainly linked to political and social
      aspects of human society. Left column presents names more linked to socialism
      and right column those more linked to capitalism; their polarization opinion
      $\Delta \mu = \mu - \mu_0$ is shown for the case of two fixed red nodes
({\it socialism, communism}) and two blue nodes
({\it capitalism, imperialism}) (OP2) for $10^3$ realisations and $10^5$ realisations: 
      $\mu_0=0.243$ and $\mu_0(L)=0.146$.
    The values of $\Delta \mu(L)$ are discussed in Section 6.}
\label{tab3}
\end{table}

  From Table~\ref{tab3} we note that 7 from 12 political leaders in right column
  have exactly the same $\Delta \mu = 0.007$ value. We attribute this to the fact that all of them
  are presidents of USA that probably is at the origin of this feature.
  Other 3 USA presidents (Roosevelt, Kennedy, Carter) have different $\Delta \mu$ values
  that we relate to extraordinary events during their time slot in office while Carter has
  $\Delta \mu$ value being not so different from 7 above presidents.

  In Tables~\ref{tab1},~\ref{tab2},~\ref{tab3} there are data
  for $\Delta \mu(L)$ obtained with a significantly higher number of
  realisations $N_r$. We discuss this data in Section 6.
  
  \begin{table}[h!]
    \centering
    \begin{tabular}{|c|c|l|c|l|}
      \hline
      $i$	&	negative $\mu$	&	Name	&	positive $\mu$ & Name	\\
    \hline
  1	&	-1.000	&	Étienne	Clavier								&			0.572	&	Giliana	Berneri								\\
  2	&	-0.786	&	Theory	of	imperialism							&			0.572	&	Maurice	Laisant								\\
  3	&	-0.777	&	Community	capitalism								&			0.572	&	Renée	Lamberet								\\
  4	&	-0.751	&	Supercapitalism:	The	Transform...	&			0.572	&	Georges	Vincey								\\
  5	&	-0.399	&	Sustainable	capitalism								&			0.572	&	Aurelio	Chessa								\\
  6	&	-0.022	&	Li	Yong	(chancellor)							&			0.572	&	Giovanna	Berneri								\\
  7	&	-0.022	&	Cheng	Yi	(chancellor)							&			0.572	&	Pio	Turroni								\\
  8	&	-0.020	&	Yu	Di								&			0.572	&	Maurice	Fayolle								\\
  9	&	-0.020	&	Emperor	Wenzong	of	Tang						&			0.572	&	Louis	Mercier-Vega								\\
  10	&	-0.020	&	Consort	Shen								&			0.508	&	Federación	Deportiva	Obrera							\\
  11	&	-0.020	&	Wu	Shaocheng								&			0.508	&	Labour	Gathering	Party							\\
  12	&	-0.020	&	Wang	Zhixing								&			0.499	&	Oneworld	(disambiguation)								\\
  13	&	-0.020	&	Shi	Yuanzhong								&			0.498	&	One	World			1964					\\
  14	&	-0.020	&	He	Jintao								&			0.498	&	Nash	Mir			1968					\\
  15	&	-0.020	&	Wang	Yuankui								&			0.498	&	Our	World								\\
  16	&	-0.020	&	Li	Deyu								&			0.474	&	Socialist	Association								\\
  17	&	-0.020	&	Liu	Zhen								&			0.474	&	Socialista									\\
  18	&	-0.020	&	Wang	Zai								&			0.474	&	Indep.	Radical	Social	Democratic	Party					\\
  19	&	-0.020	&	Shi	Xiong								&			0.468	&	Indep.	Socialist	Workers	Party						\\
  20	&	-0.020	&	He	Hongjing								&			0.456	&	Spanner	(journal)			All-Union	Communist	Party			\\
\hline
\end{tabular}
\caption{Top 20 negative and positive values of $\mu$ of articles for the case of two fixed red nodes
({\it socialism, communism}) and two blue nodes
({\it capitalism, imperialism}) (OP2). The articles of this table belong to Fig.\ref{fig5}}
\label{tab4}
\end{table}

  It is interesting to consider what are the articles with extreme $\mu $ values from Fig.~\ref{fig5}
  for OP2 case of EN edition. These articles are listed in Table~\ref{tab4}.
  To understand the reasons for such extreme $\mu$ values we consider a few examples of such articles
  with their in-going  links. Thus {\it Étienne Clavier}, who lived in 1762-1817 and was a French
  Hellinist and magistrate, is directly pointed only by {\it capitalism} article,
  since he referred in French to capitalises at very early 1788,  four yeas before
  English usage by A.Young. This leads to extreme value $\mu = -1$.

  {\it Li Yong (chancellor)} with $\mu = -0.022$ was  an official of the Chinese Tang dynasty who lived around 800 AD.
  This article 
  has such in-going link articles  as
  {\it  Index of China-related articles, Emperor Xianzong of Tang,  Chancellor of the Tang dynasty} and
  8 more in-going links of other articles
  related to China of this period. Here we see that due to links between {\it imperialism}
  and {\it imperium, empire, emperor, imperator} our approach
  leads to such an extreme value. Thus links between similar or related words can produce somewhat artificial links
  between {\it imperialism} and a person who lived in far 8 century. But such concepts as
  {\it imperium, empire, emperor, imperator} are very ancien and explains such an influence by
  the more modern concept {\it imperialism}.

  For extreme positive $\mu$ values we consider  {\it Giliana Berneri} who lived in 1919-1998
  and was French doctor of medicine and libertian communist activist. She was also among the founders of
  French Anarchist Federation, which included {\it Maurice Laisant}. Its articles has in-going links from
  articles about 
  other people linked to socialist-communist movements (Camillo Berneri,  Berneri, Georges Vincey, Aurelio Chessa,
  Giovanna Berneri) that leads to $\mu = 0.572$. The article {\it  Maurice Laisant} is ponted only
  by articles {\it Giliana Berneri, Georges Vincey} that leads to $\mu =0.572$.
 
  From these examples we see that extreme $\mu$ values appear for articles which have a small number
  of in-going links directly coming from fixed opinion articles or by a short path from them.
  
\section{Results for Christianity vs. Islam}

\begin{figure}[H]
	\begin{center}
		\includegraphics[width=0.5\columnwidth]{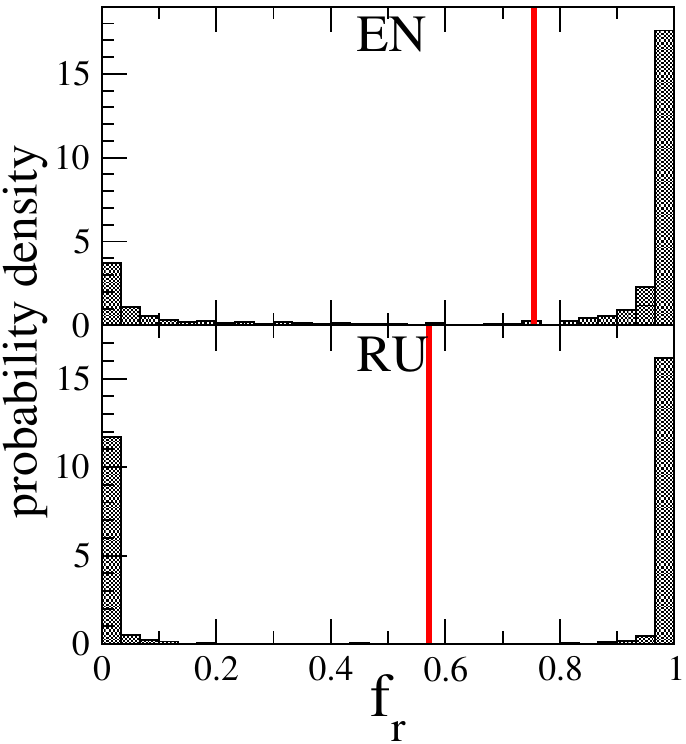}
	\end{center}
	\vglue -0.3cm
  \caption{\label{fig10}
    Same histogram as in Fig.~\ref{fig2} but for another pair of fixed nodes (articles)
  being {\it Christianity} (red) and {\it Islam} (blue) for EN (top) and RU (bottom)
      Wikipedia editions; here average opinion polarization is $\mu_0 =  0.509$ (EN),
      $0.142$ (RU) being marked by red lines; $\mu = 2f_r -1$.
  }
\end{figure}

\begin{figure}[H]
	\begin{center}
		\includegraphics[width=0.6\columnwidth]{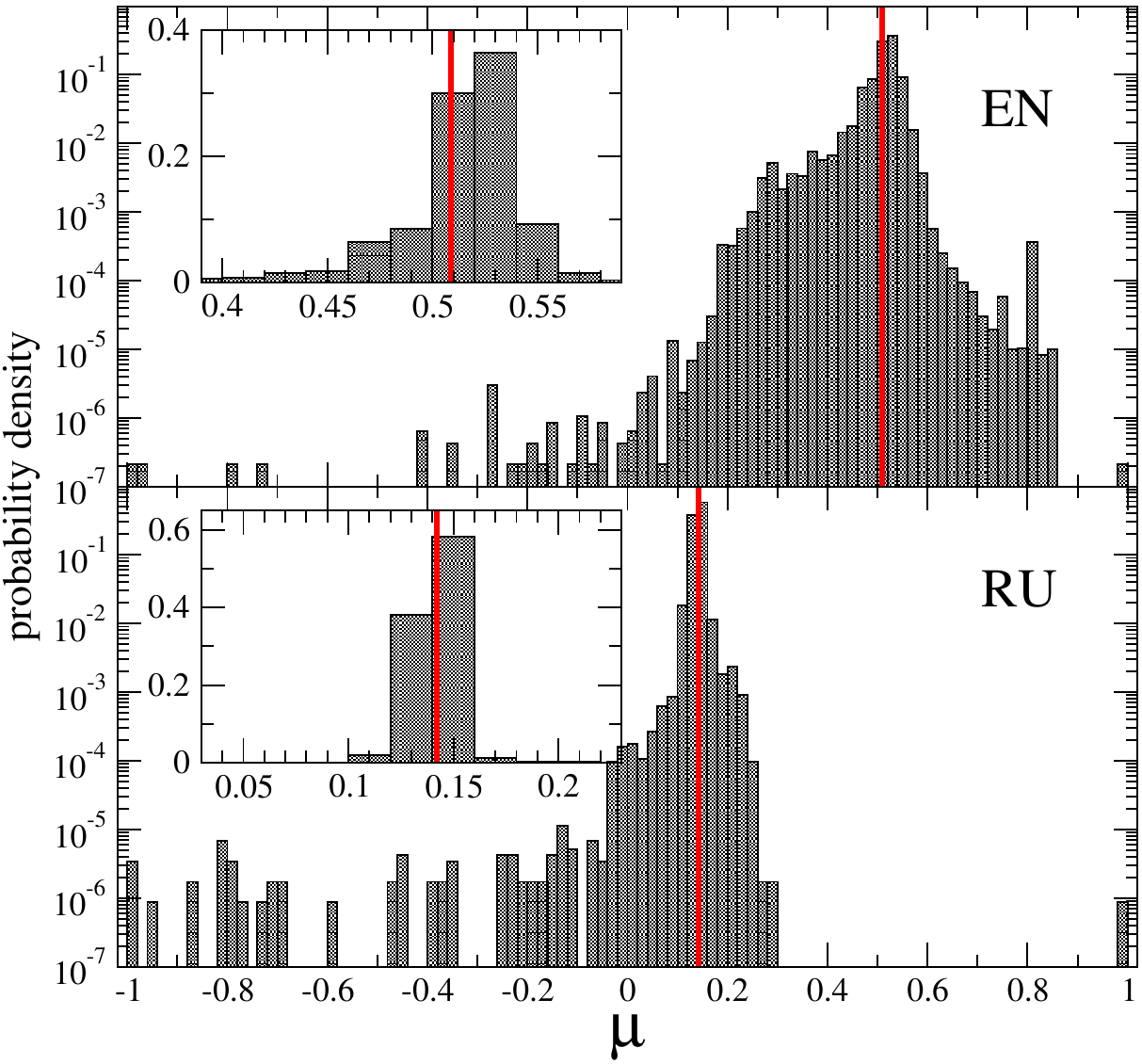}
	\end{center}
	\vglue -0.3cm
        \caption{\label{fig11} Same probability histogram as in Fig.~\ref{fig5} but
  but for another pair of fixed nodes (articles)
  being {\it Christianity} (red) and {\it Islam} (blue) for EN (top) and RU (bottom);
  red lines mark values of average global polarization opinion
   $\mu_0 =  0.509$ (EN), $0.142$ (RU)
  }
\end{figure}

Our Monte Carlo approach to opinion formation in Wikipedia networks can be
also used for another competing articles. To illustrate such another example we
consider the case of {\it Christianity} (red) and {\it Islam} (blue)
in EN and RU editions. The histograms of steady-state probability
distribution of red nodes is shown in Fig.~\ref{fig10}.
These distributions are essentially composed of two peaks at
$f_r=1$ and $f_r = 1-f_b =0$. The histograms for opinion polarization $\mu$
are shown in Fig.~\ref{fig11}. These results show
that the fraction of opinion in favor to {\it Islam}
is about by factor 3-4 higher (for $f_b, \mu_0$) in RU edition comparing to EN one.
We attribute this to a significantly higher percent of muslim population in Russia (10-12\%)
comparing to USA (1\%), UK (5\%), Canada (5\%), Australia (3\%)
(even if these percents are approximate) \cite{wikiislam}.

\begin{figure}[H]
	\begin{center}
		\includegraphics[width=0.7\columnwidth]{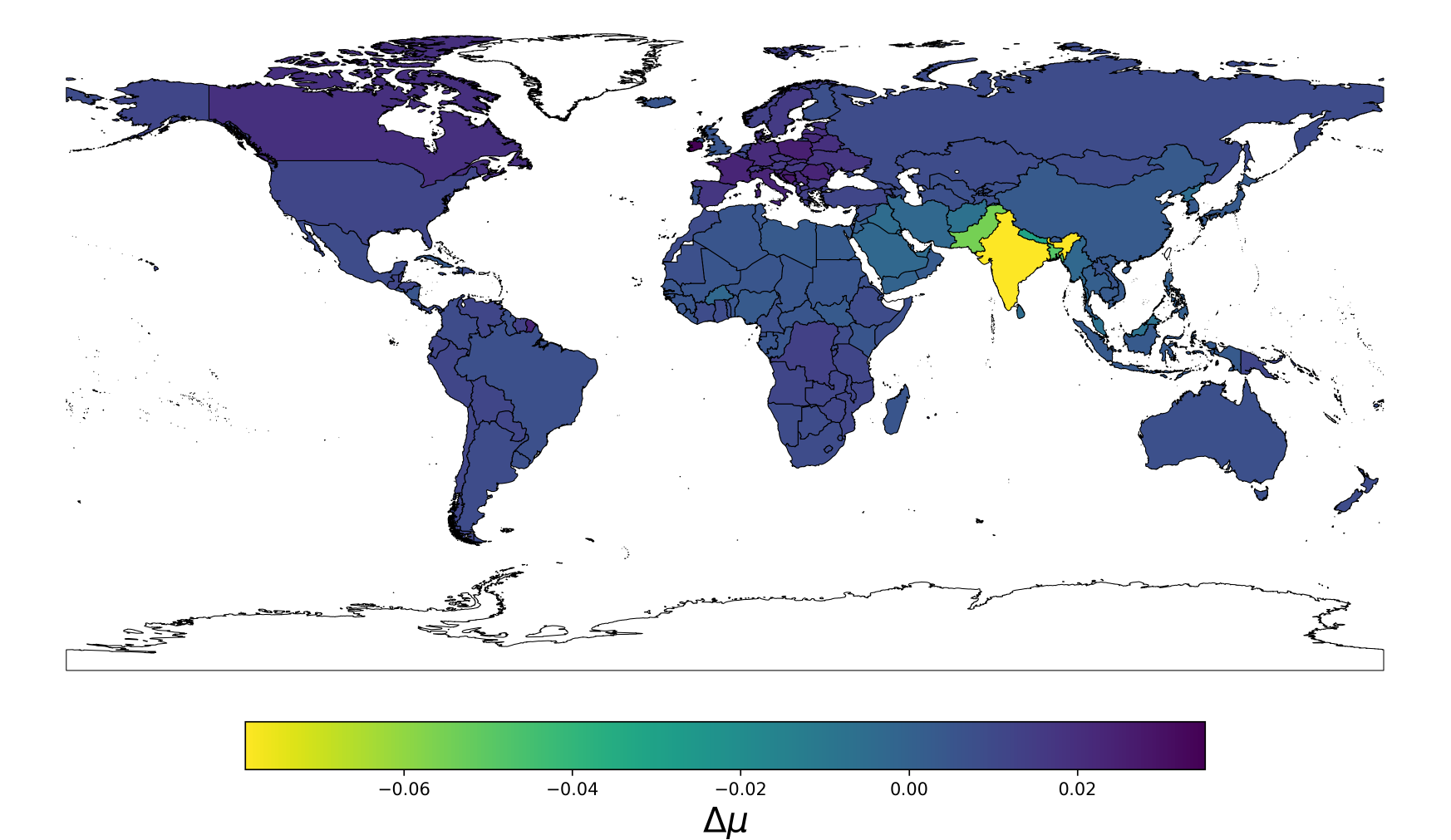}
	\end{center}
	\vglue -0.3cm
	\caption{\label{fig12}
          Geographical distribution of opinion polarization
          to {\it Christianity } ($\Delta \mu >0$)  or
          ({\it Islam}) ($\Delta \mu < 0$) expressed by $\Delta \mu$
          for English Wikipedia.
          Color legend shows the scale for $\Delta \mu$.
	}
\end{figure}

 The world map of countries characterized by their opinion polarization $\Delta \mu$
in shown in Fig.~\ref{fig12} for English Wikipedia.
The countries with extreme positive and negative opinion
polarization, expressed by $\Delta \mu$,
are in favor of {\it Christianity}:
Ireland ($0.035$), Bosnia and Herzegovina ($0.027$),
Croatia ($0.025$) and Poland ($0.025$);
and in favor of {\it Islam}: 
India ($-0.079$), Pakistan ($-0.055$), Bangladesh ($-0.049$) and Nepal ($-0.031$).
We find that the values of country $\Delta \mu$ are well correlated with
the percent of muslim population of countries $M$ taken from \cite{wikiislam}.
Thus, the correlation coefficients between $\Delta \mu $ and $M$ values
are rather high:  $0.3853$ (Pearson),
 $0.458$ (Spearman) and $0.313$ (Kendall); see definitions of coefficients at Wikipedia.

For the leading historical figures of {\it Christianity} and {\it Islam}
we obtain for EN edition $\Delta \mu$ values:
Jesus (0.019), Saint Peter (0.027), Paul the Apostle (0.029)
and Muhammad (-0.005), Ali (-0.005), Abu Nakr (-0.005).

For Russia edition for the same articles we have:
Jesus (0.00195), Saint Peter (0.00195), Paul the Apostle (0.00195)
and Muhammad (-0.006), Ali (-0.006), Abu Nakr (no such article in 2017).

We consider that these results qualitatively correspond to a natural expectation
of  opinion preference being more on the side of {\it Christianity}
for Jesus, Saint Peter, Paul the Apostle and
on the side of {\it Islam}
for Islam for  Muhammad, Ali, Abu Nakr.
This confirms the validity of our approach
for opinion formation on Wikipedia networks.

Thus the outcomes of this Section confirm that our INOF model
leads to reliable results.

\section{Results for Democratic Party vs. Republican Party in USA}

As an another example of competition
between two opinions we inside the case of two articles
in EN edition:
{\it Republican Party (United States)} (red)
and
{\it Democratic Party (United States)} (blue).
In this case the histogram analogous to one of Fig.~\ref{fig2}
is still essentially composed of 2 peaks of different heights at
$f_r =1$ and $f_r=1-f_b =0$,
$\mu_0 = -0.435$ with red and blue fractions being
$f_r = (1+\mu_0)/2=0.282$ and $f_b=1-f_r = 0.718$.
The article {\it United States} has $\mu _{US}= -0.452$
with $\Delta \mu_{US} =-0.017$.
Thus the EN edition is  significantly more favorable
for Democratic Party. 

In global, on the basis of obtained results for directed networks of 6 Wikipedia editions
we conclude that our INOF model gives reliable understated for confrontations
of two opposite opinions in such systems.

\section{Statistical features of INOF model}

For a given edition the value of average opinion polarization $\mu_0$
is determined from $N_r=1000$ realisations and
$N$ spins of a given realisation (we mark this as a slot 1 discussed in
previous Sections). Thus e.g. for EN edition
$\mu_0$ is obtained from summation over approximately $5 \times 10^9$
spin orientations while $\mu$ values of articles are obtained from 1000
spins. Thus one would expect that the values of global polarization
$\mu_0$ and polarization of individual article $\mu_i$ are
statistically very stable. However, when we perform
a comparison with another slot 2 with other $N_r=1000$ random
pathways we obtain a notable change of $\mu_0$ and $\mu_i$ values.
At the same time, the fractions of white nodes in the steady-state, related to isolated communities,
remain the same for different slots. 
Also the extreme values of $\mu$, as those in Table~\ref{tab4},
have little changes or nothing for different slots
in a difference from the articles in the
main part of probability distribution.
We attribute this to the fact that such extreme
articles have short links to the fixed nodes
and hence are only weakly affected by pathway realisations.

\begin{figure}[H]
	\begin{center}
		\includegraphics[width=0.5\columnwidth]{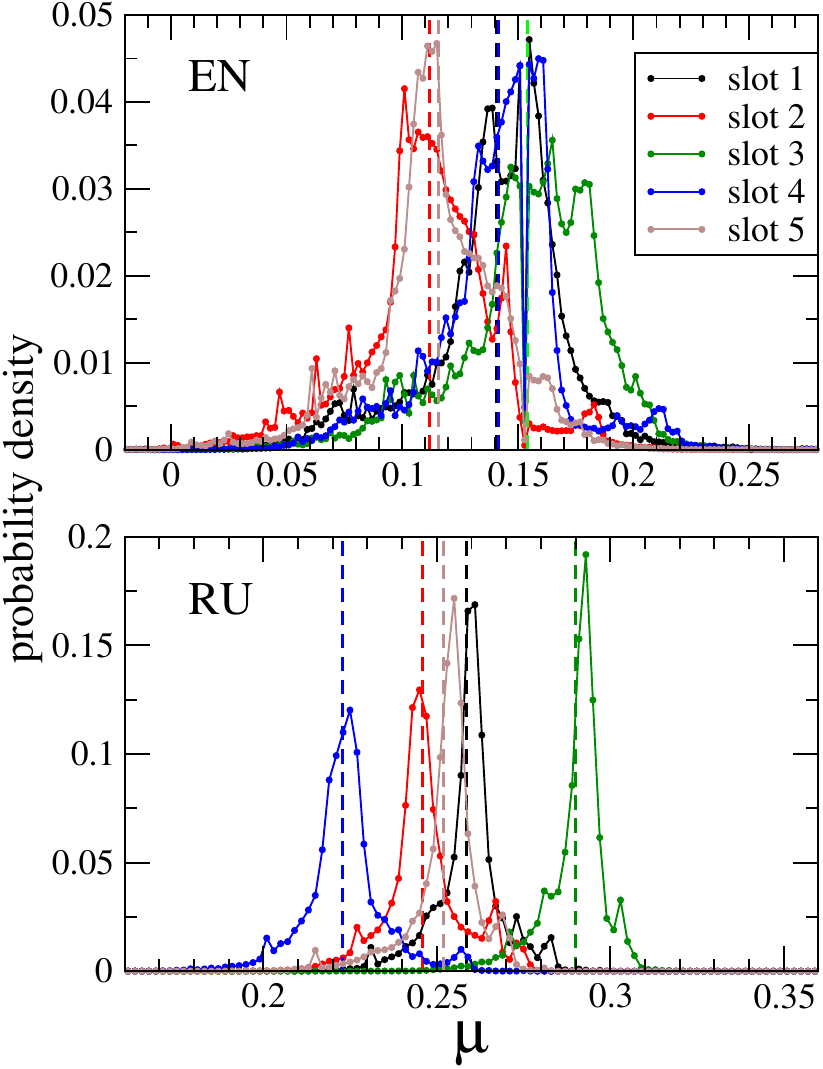}
	\end{center}
	\vglue -0.3cm
  \caption{\label{fig13}
  Probability density of the average opinion polarization value $\mu$ for the English
  edition for OP2 (top panel) and for the Russian edition for OP2 (bottom panel).
  The five slots of the model are represented by curves of different colors, with 1000 realisations 
  per slot for EN and 2000 for RU. The slot 1 discussed in previous Sections has black color.
  The bin width in $\mu$ is $10^{-3}$, and $\mu_0$ values of slots are represented by dashed vertical 
  lines corresponding to the same color as the distribution.
  }
\end{figure}

As an example, we show in Fig.~\ref{fig13} the probability
distributions for 5 random slots with $N_r=1000$ for  EN edition
and 5 slots with $N=2000$ for RU edition. There is a visible  modification
of form of distribution. The values of 5 $\mu_0$ are
approximately varying in a range of 25\%-30\% for RU and EN editions
comparing to the average of these 5 values of $\mu_0$.

\begin{figure}[H]
	\begin{center}
		\includegraphics[width=0.7\columnwidth]{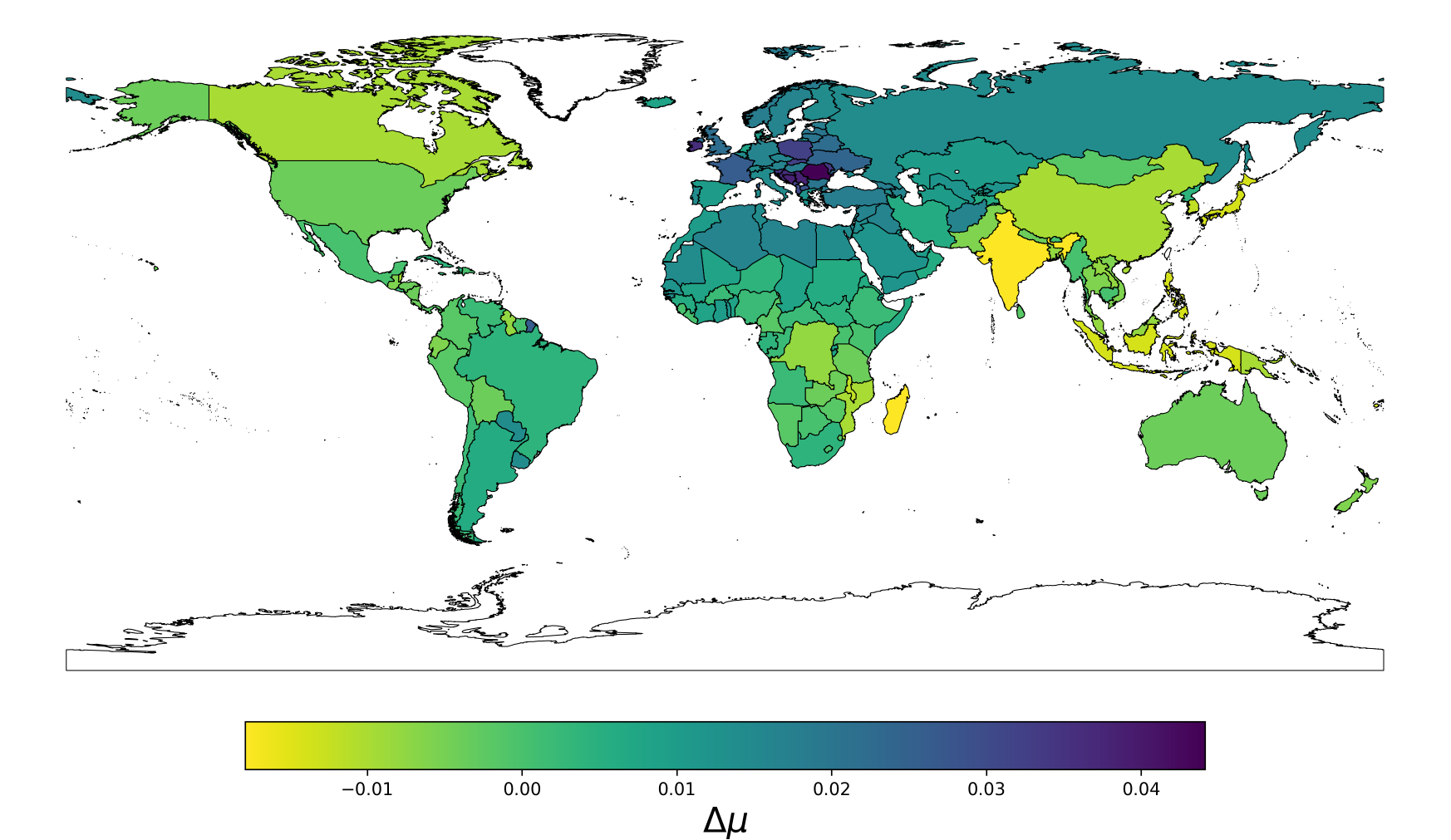}
	\end{center}
	\vglue -0.3cm
        \caption{\label{fig14}
         Same as in Fig.~\ref{fig9} for the slot 5 of EN edition marked by light brown color in Fig.~\ref{fig13}.
  }
\end{figure}

The effect of $\mu$ variations for specific articles of 194 world countries  is shown in Fig.~\ref{fig14}
as the world map of countries for slot 5 to be compared with the result of Fig.~\ref{fig9}
for slot 1. We see that the individual values $\Delta \mu = \mu - \mu_0$  of countries are changed in these two Figures
but the global features of opinion polarization remain similar.

To characterize the similarity between $\mu$ values in presented 5 slots of EN and RU editions in Fig.~\ref{fig13}
we compute correlators between $\mu$ values of these 5 slots. There are 10 different correlators from
5 slots of EN (and 10 for 5 of RU). These 10 correlators have similar values $C$ and due to that
we give here only their average value and the standard deviation obtained from these 10 correlators.
Thus for correlators of only articles of 194 countries  
we obtain $C= 0.781 \pm 0.038$ (Spearman); $0.795 \pm 0.042$ (Pearson); $0.623 \pm 0.038$ (Kendall)
for OP2 case of EN edition.
If we compute these 10 correlators for all $N$ articles of OP2 case of EN edition then we find 
$C= 0.851 \pm 0.031$ (Spearman); $ 0.949 \pm 0.009$ (Pearson); $ 0.696 \pm 0.036$ (Kendall),
thus for all articles the correlators are even higher. The definitions of the used
three correlators of Spearman, Pearson, Kendall can be find in Wikipedia.

For  5 slots of RU edition of Fig.~\ref{fig13} with all articles we have similar values of
correlators being
$C= 0.860 \pm 0.014$ (Spearman); $0.998 \pm 0.001$ (Pearson); $0.719 \pm 0.017 $ (Kendall).

Thus the correlator analysis shows that different slots have highly correlated $\mu$ values
but the fluctuations of $\mu$ values from slot to slot are still significant for
number of realisations $N_r=1000$ used in previous Sections.

\begin{figure}[H]
	\begin{center}
		\includegraphics[width=0.5\columnwidth]{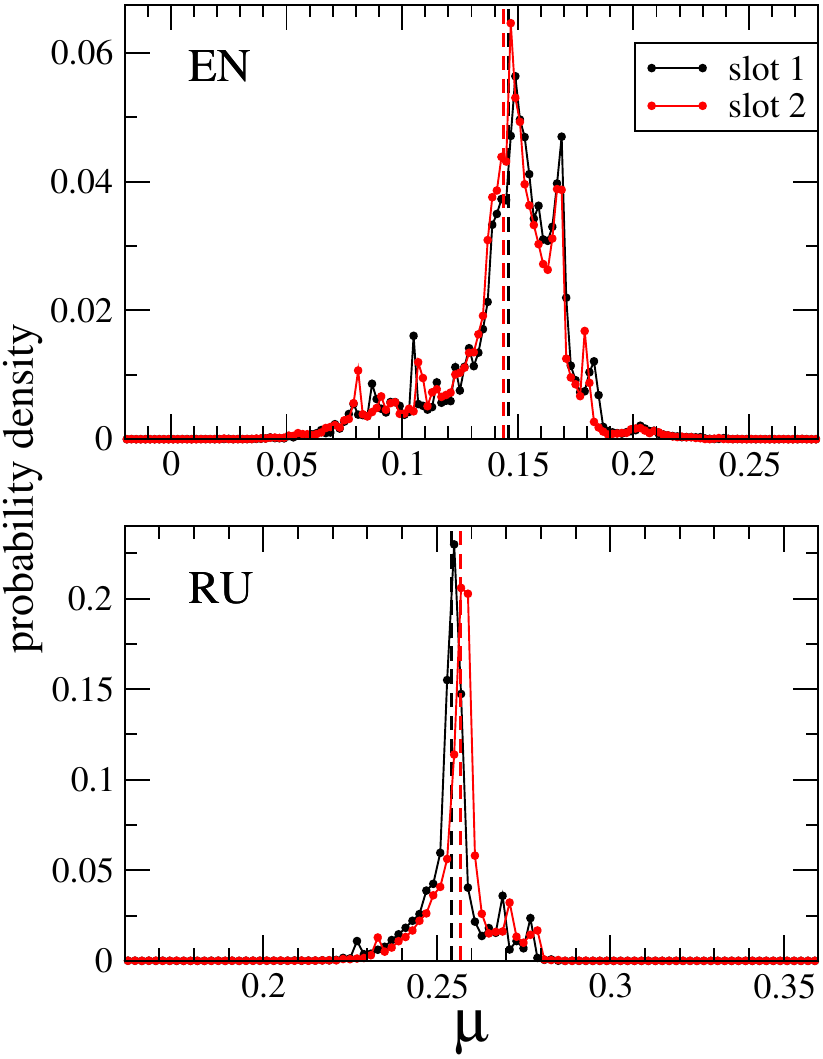}
	\end{center}
	\vglue -0.3cm
	\caption{\label{fig15} 
  Probability density of the average opinion polarization value $\mu$ for the English
  edition for OP2 (top panel) and for the Russian edition for OP2 (bottom panel).
  The two slots of the model are represented by black and red curves, with $10^5$ realisations 
  per slot for EN and RU.
  The bin width in $\mu$ is $5.10^{-4}$, and $\mu_0$ values of slots are represented by dashed vertical 
  lines corresponding to the same color as the distribution.  The data of slot 1  are marked by black color,
  they are used in Tables~\ref{tab1},~\ref{tab2},~\ref{tab3}.
	}
\end{figure}

With the aim to reduce these fluctuations of opinion polarization
we significantly increased the number realisation going up to
$N_r=10^5$. This allows to obtain a significant reduction of
fluctuations of $\mu$ values of individual articles
as it is shown in Fig.~\ref{fig15} for EN and RU editions of OP2 case.
We note that a run with $10^5$ realisations for EN edition
takes 5 days of CPU time on 40 core processor.

\begin{figure}[H]
	\begin{center}
		\includegraphics[width=0.7\columnwidth]{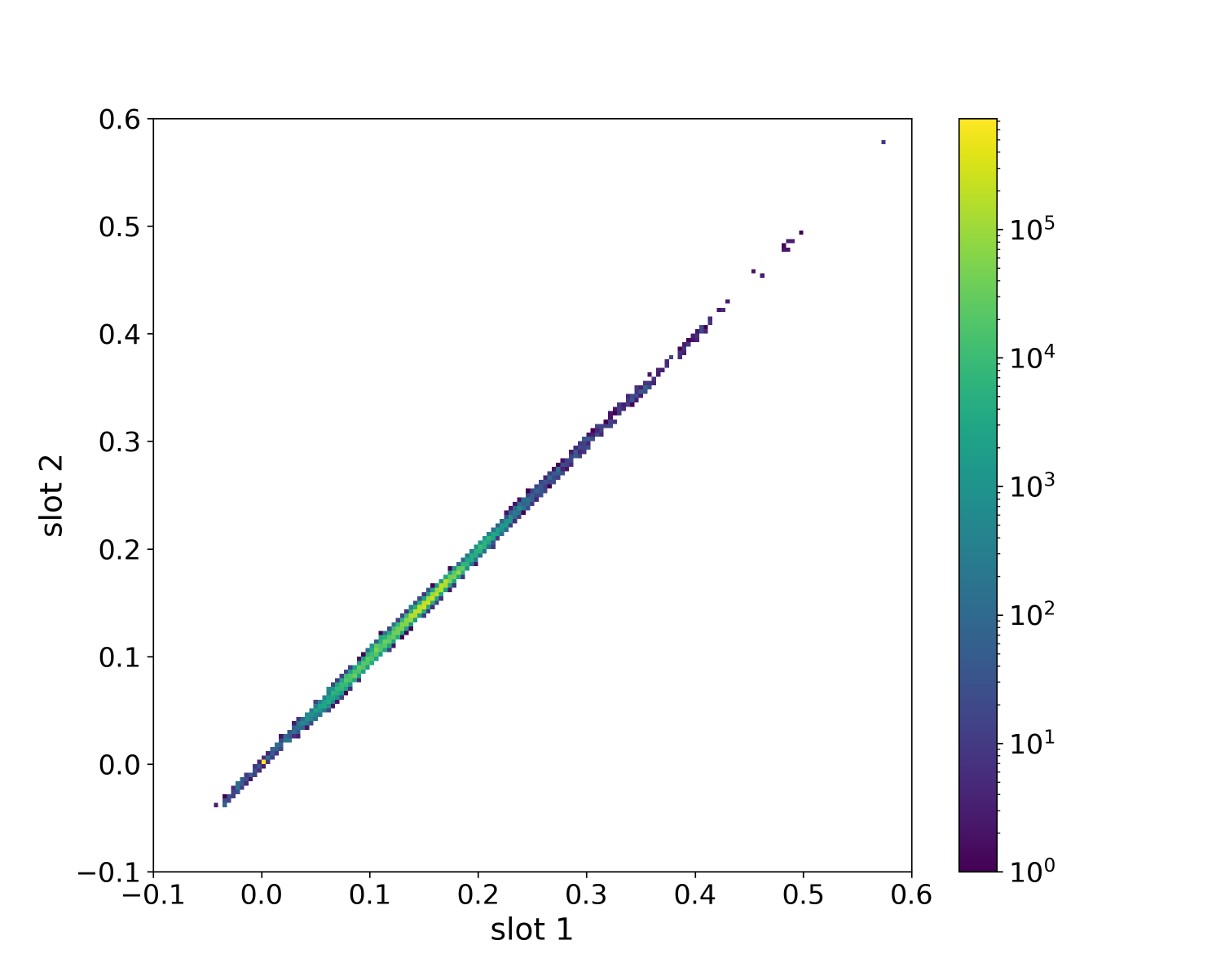}
	\end{center}
	\vglue -0.3cm
	\caption{\label{fig16} 
          Density distribution of number of articles
          in the plane of $(\mu_1,\mu_2)$
          values for the INOF model across two slots with $10^5$ realisations each. 
  Each article has a $\mu_{1,2}$ value, given in axes, for {\it slot 1} and {\it slot 2}, 
  and the number of articles in this plane is represented by a 
  color scale in the density distribution using a logarithmic scale. 
  White indicates regions without articles.
	}
\end{figure}

To illustrate a difference between two slots we show the density of articles
in a plane of  their $\mu$ values, averaged over $N_r=10^{5}$ realisations,
for {\it slot 1} and {\it slot 2} of EN edition (OP2 case)
shown in Fig.~\ref{fig16}. The width of the distribution characterizes the fluctuations of $\mu$ values
being maximal near the global average $\mu_0$ values being $0.146$, for  {\it slot 1}
where the density of articles is the highest. For articles at the extreme $\mu$ values the fluctuations  are
reduced that we attribute to short pathways between these articles and the fixed ones of red or blue color.

\begin{figure}[H]
	\begin{center}
		\includegraphics[width=0.5\columnwidth]{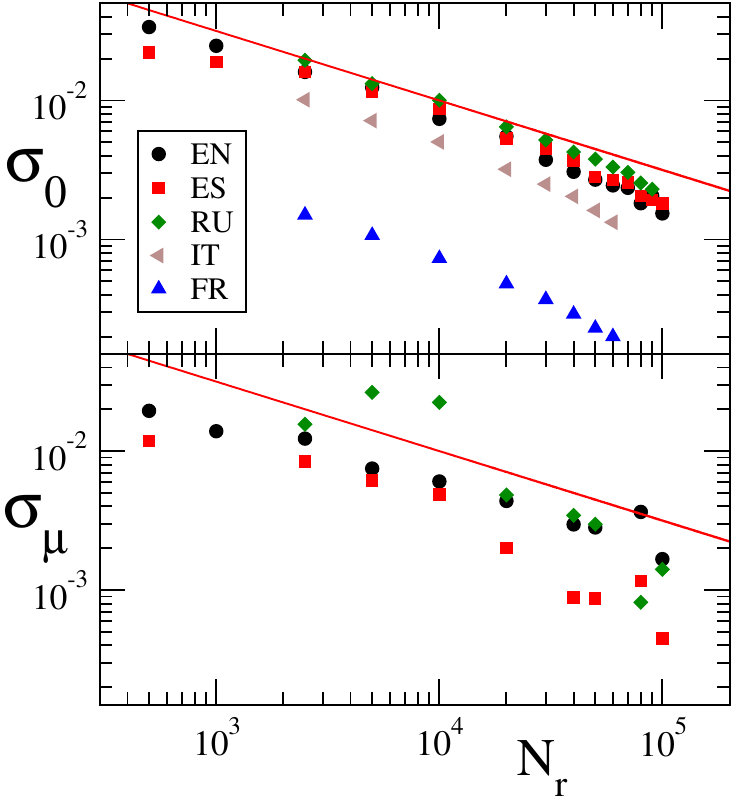}
	\end{center}
	\vglue -0.3cm
	\caption{\label{fig17} 
          Top panel shows the average variation $\sigma_0$ of $\mu_0$ for different slots 
          as a function of the number of realisations per slot ($N_r$).
          The power law fit for EN, ES, RU, FR and IT languages have exponents  $\eta$ being
          $-0.58$, $-0.5$, $-0.58$, $-0.62$ and $-0,6$ respectively.
           Bottom panel represents
          the average dispersion of individual article polarization $\sigma_\mu$ vs. $N_r$ 
          The power law fit for EN, ES and RU languages have exponents $\eta_{\mu}$ being
          $-0.42$, $-0.62$ and $-0.0.85$ respectively.
          Red line in both panels illustrates the power law with exponent $-1/2$
          with $\sigma_{0,\mu}=1/\sqrt{N_r}$ .
          The number of slots used to compute $\sigma_{0,mu}$ varies from $40$ for $N_r=500,2500$
          to $2$ for $N_r=10^5$. 
	}
\end{figure}

To obtain a quantitative characterization of
$\mu_0$ fluctuations and their dependence on the $N_r$ value, we define
an average variation of $\sigma_{0}$ as
$\sigma_0 = \sqrt{\frac{1}{N_s} \sum_{j=1}^{N_s} (\mu_{0,j}-\langle\mu_0\rangle)^2}$,
where $\mu_{0,j}$ is the average polarization of slot $j$, $N_s$ is the number of slots,
and $\langle\mu_0\rangle$ is the average $\mu_0$ for $N_s$ slots.
We also define the average dispersion of individual article polarization $\sigma_\mu$ as
$\sigma_\mu=\sqrt{\frac{1}{N} \sum_{i=1}^N(\mu_1(i) - \mu_2(i))^2}$, where two different
slots ($1$ and $2$) are compared in each of the $N$ articles, and
the result is averaged over different slot pairs.

The dependences of $\sigma_0$ and $\sigma_{\mu}$ on $N_r$ are presented in Fig.~\ref{fig17}
for  Wikipedia editions. For $\sigma_0$ the dependence on $N_r$  is well described by an expression
$\sigma_{\mu} \approx B/{N_r}^{1/2}$ with $B \approx 1.5$
(for FR case $B \approx 0.23$, that can be attributed to
that it has $\mu_0$ being very close to 1).
The fits of decay exponent $\eta$ give $\eta = -0.57 \pm 0.06$
from 5 editions; this value is very close to $\eta=1/2$ corresponding
to the inverse square root decay. For $\sigma_{\mu}$ the exponent $\eta_{\mu}$
is also close to $1/2$ for EN, ES editions while for RU edition flustuations
with $N_r$ are too high to get a reliable value of $\eta_{\mu}$.
At present we have no theoretical
explication for the exponent $\eta$ being close to $1/2$ for the main part of editions.

\begin{figure}[H]
	\begin{center}
		\includegraphics[width=0.7\columnwidth]{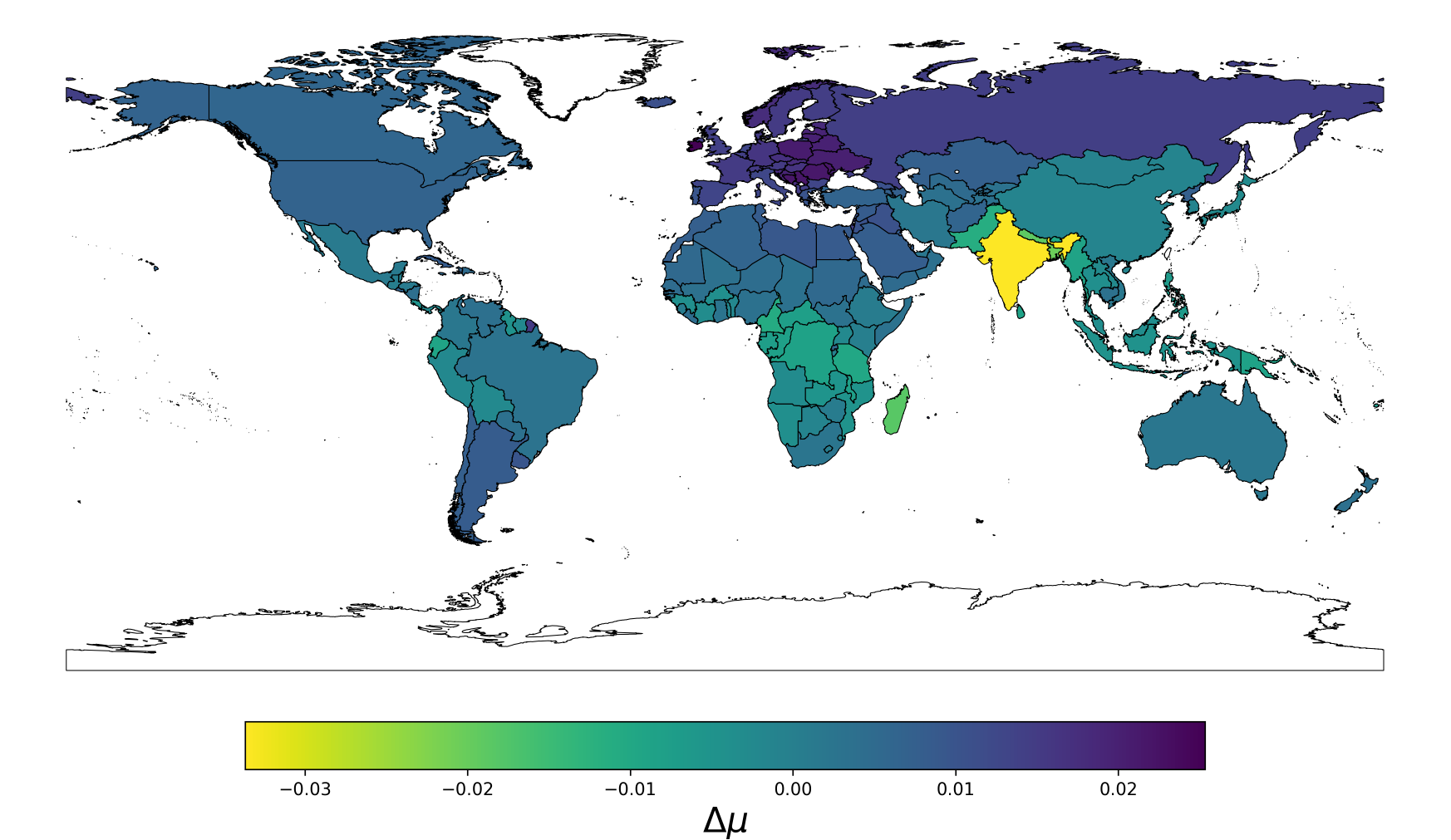}
	\end{center}
	\vglue -0.3cm
	\caption{\label{fig18} 
          Geographical distribution of opinion polarization preference
          to {\it socialism, communism} ($\Delta \mu >0$) or
          ({\it capitalism, imperialism}) ($\Delta \mu <0$) (OP2) expressed by $\Delta \mu$
          for English Wikipedia and long run of $10^5$ realisations ({\it slot 1} in Fig.~\ref{fig15}).
          Color bar shows the scale for $\Delta \mu$.
	}
\end{figure}

In Tables~\ref{tab1},~\ref{tab2},~\ref{tab3} for specific articles
we compare the values of $\Delta \mu_0$ obtained with $N_r=10^3$ and $N_r=10^5$
realisations. Practically for all articles presented in these tables
the difference of values is only in the third digit that approximately corresponds
to $\sigma_{\mu}$ standard deviation from Fig.~\ref{fig17}. Thus the small values of $\Delta \mu$
should be taken with a caution. As an example of changes in $\Delta \mu$ at higher statistics of $N_r$
we may note e.g. {\it United States}, {\it Brazil}, {\it Turkey} in Table~\ref{tab2}
that are getting positive $\Delta \mu(L)$ values at higher statistics. But still the changes of $\Delta \mu$
are in the third digit. In Table~\ref{tab3} the high number of $M_r=10^5$ moves
{\it Mao Zedong} to positive $\Delta \mu(L)$ value, from the capitalistic side of this Table 
all politicians with negative $\Delta \mu$ at $N_r=10^3$ are moved to positive
$\Delta \mu(L) $ values at $N_r = 10^5$; but still their $\Delta \mu(L) $ values remain by
a factor 3 smaller compared to the case of politicians with the socialistic orientation in the left column.

Finally in Fig.~\ref{fig18} we show the opinion polarization $\Delta \mu$ for world countries
for OP2 case of EN edition obtained with $N_r=10^5$ realisations ({\it slot 1} in Fig.~\ref{fig15}).
There is a clear dominance of socialistic orientation
for a main number of countries especially in Europe and Russia.
The global feature of this world map are similar to those shown in Figs.~\ref{fig9},~\ref{fig14}
obtained with $N_r=10^3$. But it is clear that results of Fig.~\ref{fig18} are
much more stable in respect to fluctuations.

An interested reader can find
the map of world countries opinion polarization for all 6 Wikipedia editions 
at high $N_r =10^5$ in \cite{ourpage}. For all articles of these 6 editions
the opinion polarization $\mu$ values for OP2 case are also available at \cite{ourpage}.

\section{Discussion and conclusion}

We developed  the Ising Network Opinion Formation (INOF) model
and applied it to analysis of opinion formation in Wikipedia 
networks of 6 language editions of year 2017. In this model Ising spins with fixed
opposite directions present certain fixed opinions, red or blue, of selected network nodes.
All other nodes have initially zero spin of white opinion.
Then the Monte Carlo step procedure determines inversion of spins
determined by their in-going links until a steady-state 
polarization of all network spins is reached.
This allows to determine the global opinion preference of the whole network
as well as opinion polarization of individual nodes.

We mainly considered the confrontation of {\it capitalism, imperialism} and
{\it socialism, communism}. We  find that for 6 Wikipedia editions (EN, DE, ES, FR, IT, RU)
a majority opinion is in favor of {\it socialism, communism}.
The variations of opinion preferences for the world countries, political leaders
and other Wikipedia articles are determined being in a good
agreement with simple heuristic expectations. We give also
arguments for certain deviations from such expectations.

In addition we consider the opinion formation given by
interactions between {\it Christianity} and {\it Islam}
for EN and RU editions. The INOF model naturally gives a
significant preference for {\it Christianity} in EN Wikipedia
while the preference of RU Wikipedia is significantly more balanced.
The INOF model determins the preference balance for the world countries
which has a high correlation coefficient
with the muslim population of countries for EN and RU editions.

We also consider the competition of US Democratic and Republican Parties in EN Wikipedia of 2017.
The global opinion preference is found to be significantly in favor of Democrats.

We note that the INOF model may have some uncertain
situations, like for example the case of article {\it China}
which has in-going direct links from {\it capitalism, imperialism}
and from {\it socialism, communism}. However, in great majority of
studied cases the model gives good realistic opinion preferences.

On the basis of obtained results we expect that the proposed INOF model
will find various applications for opinion formation
in numerous directed networks.

%

\begin{acknowledgments}
The authors acknowledge support from the grant
 ANR France project
NANOX $N^\circ$ ANR-17-EURE-0009 in the framework of 
the Programme Investissements d'Avenir (project MTDINA).

We thank K.M. Frahm for useful discussions.
\end{acknowledgments}




\end{document}